\documentclass[aip,reprint]{revtex4-1}

\draft 
\usepackage{amsmath}
\usepackage{graphicx}
\usepackage{xcolor}
\usepackage{subfigure}
\usepackage{caption}
\usepackage[colorinlistoftodos]{todonotes}

\usepackage{xr}

\begin{document}


\title{Phase behaviour of hard cylinders
} 


\author{Joyce T. Lopes}
\email{joycejtl@gmail.com}
\affiliation{Universidade Estadual de Campinas, Faculdade de Engenharia Química, Departamento de Engenharia de Sistemas Químicos}

\author{Flavio Romano} 
\email{flavio.romano@unive.it}
\affiliation{Dipartimento di Scienze Molecolari e Nanosistemi, 
Universit\`{a} Ca' Foscari di Venezia
Campus Scientifico, Edificio Alfa,
via Torino 155, 30170 Venezia Mestre, Italy}
\affiliation{European Centre for Living Technology (ECLT)
Ca' Bottacin, 3911 Dorsoduro Calle Crosera, 
30123 Venice, Italy}

\author{Eric Grelet}
\email{eric.grelet@crpp.cnrs.fr}
\affiliation{Université de Bordeaux, CNRS, Centre de Recherche Paul-Pascal, 115 Avenue Schweitzer, 33600 Pessac, France}

\author{Lu\'is F. M. Franco}
\email{lmfranco@unicamp.br}
\affiliation{Universidade Estadual de Campinas, Faculdade de Engenharia Química, Departamento de Engenharia de Sistemas Químicos}

\author{Achille Giacometti} 
\email{achille.giacometti@unive.it}
\affiliation{Dipartimento di Scienze Molecolari e Nanosistemi, 
Universit\`{a} Ca' Foscari di Venezia
Campus Scientifico, Edificio Alfa,
via Torino 155, 30170 Venezia Mestre, Italy}
\affiliation{European Centre for Living Technology (ECLT)
Ca' Bottacin, 3911 Dorsoduro Calle Crosera, 
30123 Venice, Italy}


\date{\today}

\begin{abstract}
Using isobaric Monte Carlo simulations, we map out the entire phase diagram of a system of hard cylindrical particles of length $L$ and diameter $D$, using an improved algorithm to identify the overlap condition between two cylinders. Both the prolate $L/D>1$ and the oblate $L/D<1$ phase diagrams are reported with no solution of continuity. In the prolate $L/D>1$ case, we find intermediate nematic \textrm{N} and smectic \textrm{SmA} phases in addition to a low density isotropic \textrm{I} and a high density crystal \textrm{X} phase, with \textrm{I-N-SmA} and \textrm{I-SmA-X} triple points. An apparent columnar phase \textrm{C} is shown to be metastable  as in the case of spherocylinders. In the oblate $L/D<1$ case, we find stable intermediate cubatic \textrm{Cub}, nematic \textrm{N}, and columnar \textrm{C} phases with \textrm{I-N-Cub}, \textrm{N-Cub-C}, and \textrm{I-Cub-C} triple points. Comparison with previous numerical and analytical studies is discussed. The present study, accounting for the explicit cylindrical shape, paves the way to more sophisticated models with important biological applications, such as viruses and nucleosomes.
\end{abstract}


\maketitle 



\section{Introduction}
\label{sec:introduction}
After nearly one century since Onsager's pioneering prediction that orientational order can be entropically induced for elongated particles~\cite{Onsager1949}, simple models of rod-like objects continue to play a central role in the study of colloidal liquid crystals \cite{Gibaud2012,Grelet2014} and self-assembly processes \cite{Liang2017,Sung2018}.  

The simplest model for a rod-like molecule is the hard spherocylinder, an object formed by a cylinder of length $L$ capped with two hemispheres of matching diameter $D$. This shape can be obtained by rolling a sphere of \textcolor{black}{radius} $D/2$ around a segment of length $L$. The great advantage of this model, and the key to its popularity, is the simplicity of the overlap condition between two such hard spherocylinders; this condition can be cast in a simple analytical form \cite{Allen1993,Vega1994} that can be computed very efficiently. As early as 1997, \citeauthor{Bolhuis1997} \cite{Bolhuis1997} performed a remarkably detailed study of the phase diagram of this model that is now reckoned as a classic reference in the field. Other similar shapes have also been proposed in the literature, including hard ellipsoids \cite{Frenkel1985}, hard helices \cite{Frezza2013}, and hard dumbbells \cite{Milinkovic2013}.

However, there are physically relevant objects whose shape cannot be represented as  hard spherocylinders but rather as hard cylinders. Examples include biologically relevant cases such as viruses \cite{Bernal1941,Wen1989,Grelet2014} and nucleosomes \cite{Leforestier2008,Livolant2012a}. Hard cylinders of length $L$ and diameter $D$ have also the additional advantage of having a natural oblate limit $L/D<1$, approaching a disk for $L/D \rightarrow 0$, as well as the prolate limit $L/D>1$ (rod). This is not the case of hard spherocylinders where the oblate limit is obtained by resorting to a slightly modified model \cite{Bolhuis1997}. By contrast the overlap condition between two cylinders is significantly more evolved with respect to the spherocylinder case. This notwithstanding, and given the similarity in shape, one might rightfully wonder what are the differences, if any, in the two phase diagrams. For instance, the phase diagram of hard ellipsoids \cite{Odriozola2012} is different from the phase diagram of hard spherocylinders, in spite of the significant similarities in their shapes. This issue goes far beyond a simple academic problem in view of the strong propensity of nucleosomes \cite{Leforestier2008,Livolant2012a} and filamentous viruses \cite{Grelet2008,Grelet2016} to form a columnar phase, whose existence in the phase diagram of spherocylinders has been ruled out by recent detailed numerical simulations\cite{Dussi18} for prolate particles even if it were predicted theoretically\cite{Wensink2009} for oblate ones.

The aim of the present paper is to tackle this issue by performing a detailed analysis of the phase diagram of hard cylinders both in the prolate ($L/D>1$) and in the oblate ($L/D<1$) cases. While simulations of hard cylinders have been performed in the past\cite{Allen1993,Blaak1999,Orellana2018}, to the best of our knowledge, our study is the first one providing the complete phase diagram. For this purpose, we perform isobaric Monte Carlo simulations of a system of hard cylinders in a wide range of aspect ratios $L/D$ and volume fractions, using an efficient method for the overlap test that compares well with existing ones \cite{Allen1993,Blaak1999,Orellana2018}. The algorithm, inspired by Ref.~\onlinecite{Orellana2018}, is described in the Appendix. By monitoring the appropriate order parameters and correlation functions, we provide the corresponding phase diagram in the $L/D$ volume fraction plane and compare it with the corresponding phase diagram of the hard spherocylinders \cite{Bolhuis1997}. Particular care has been devoted in avoiding possible finite size effects along the lines of a recent similar analysis for spherocylinders \cite{Dussi18}.

The outline of the paper is as follows. In Section \ref{sec:simulations} we described the details of our numerical approach, as well as the arsenal of tools (order parameters and correlation functions) useful to identify all different phases. Some technical details have been confined in Appendix \ref{sec:algorithm}. Section \ref{sec:results} will report the main results of the present study with additional Figures and Tables reported in Section \ref{sec:supplementary}. Finally, Section \ref{sec:conclusions} reports the key messages of this study and some interesting perspectives for the future. 

\section{Simulations}
\label{sec:simulations}
\subsection{Monte Carlo simulations}
Our particles consist of $N$ cylinders/disks of height $L$, diameter $D$, and whose orientations are identified by a unit vector $\widehat{\mathbf{u}}$, as shown in Figure \ref{fig:confcy} (a). Pressures are measured in reduced units $P^{*}=P v_{\mathrm{HC}}/k_B T$, and the density $\rho = N/V$  is represented by the volume fraction $\eta \equiv N v_{\mathrm{HC}} /V$, where $v_{\mathrm{HC}}=L \pi D^2$/4 is the volume of a hard cylinder (HC). We then performed isobaric (NPT) Monte Carlo (MC) simulations at different aspect ratios $L/D$ both for rods ($L/D>1$) and for disks ($L/D<1$). All simulations were organised in cycles (MC steps), each consisting on average of $1000$ attempts to translate and rotate a randomly selected particle, and one attempt to change the volume of the simulation box. In all cases, we have performed compression runs starting at low pressure in the isotropic phase, and an expansion run starting from a close-packed solid configuration at high pressure. 
Each system was first equilibrated using $\approx 5.45 \times 10^6$
MC steps, with additional production runs of $1.5\times 10^5$ steps. The typical number of particle was $N \approx 1000$, but different numbers were used depending on the aspect ratio, as detailed in Tables~\ref{rodsnld} and ~\ref{disksnld}.
In the case of disks, the number of particles $N$ was adjusted depending on $L/D$ to keep the simulation box roughly cubic. \textcolor{black}{In our NPT simulations, we have used floppy (i.e. shape-adapting) rectangular computational box, where one axis was randomly selected and its length was allowed to change, with periodic boundary conditions to obtain an isotropic pressure \cite{Bolhuis1997} in the prolate $L/D>1$ case, and simple uniform volume move with cubic periodic boundary conditions in the oblate $L/D<1$ case. In some specific cases, we have also extended the computational box along the main axis of the cylinders to minimise finite size effects \cite{Dussi18}.  }

\subsection{Overlap of hard cylinders}\label{hcovlp}
\begin{figure}[ht!]
	\subfigure[]{\includegraphics[height=0.25\textheight]{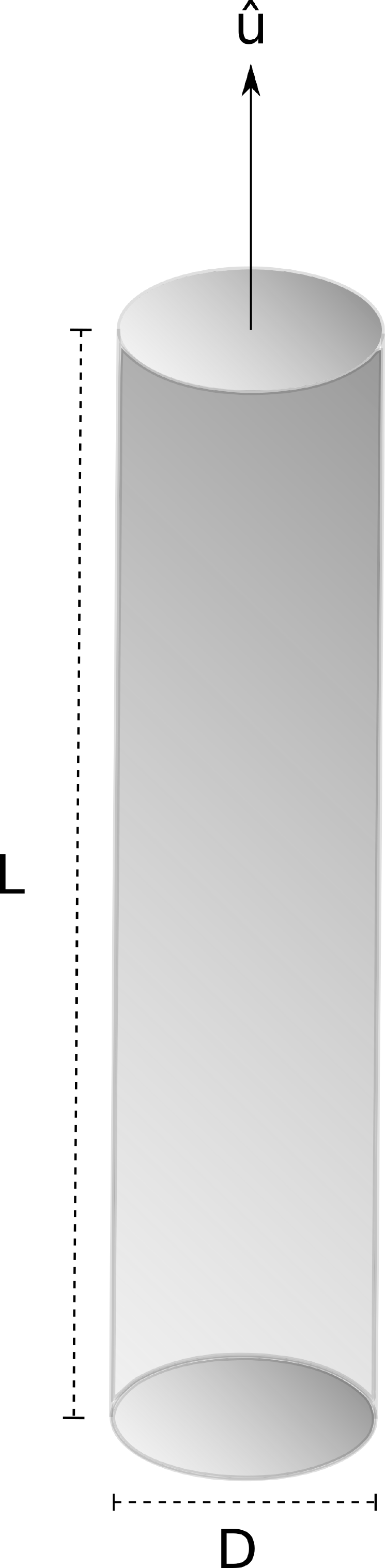}}
	  \subfigure[]{\includegraphics[width=0.2\textwidth]{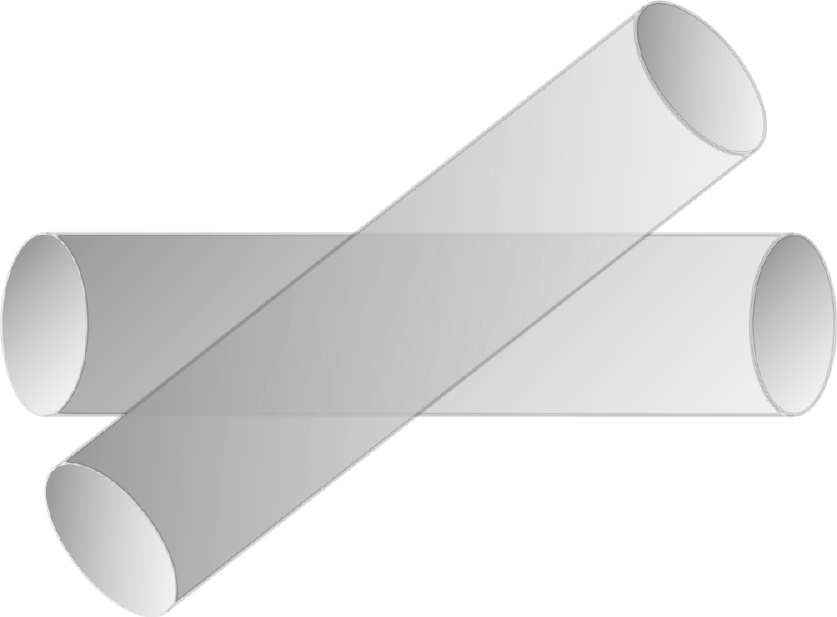}}
	  \subfigure[]{\includegraphics[width=0.2\textwidth]{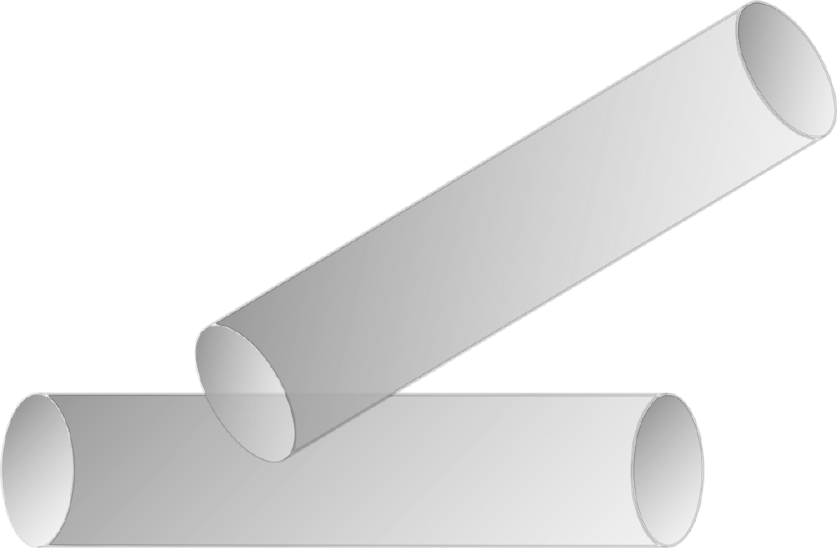}}
	  \subfigure[]{\includegraphics[width=0.15\textwidth]{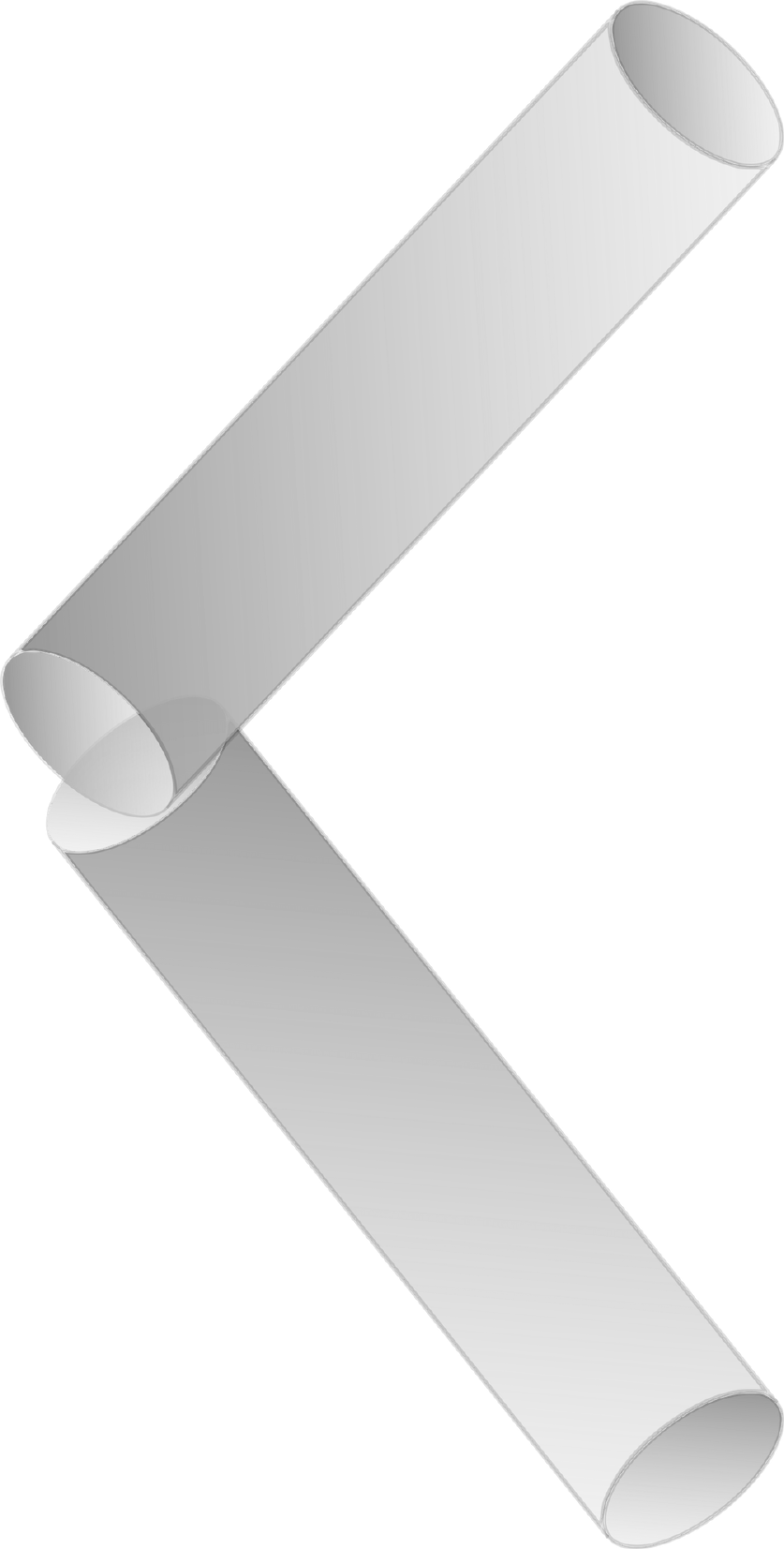}}
	\caption{(a) Our cylinder model, where $L$ is the height, $D$ is the diameter, $\widehat{\mathbf{u}}$ is the unit vector defining the orientation of the cylinder; possible overlap configurations between two cylinders (see appendix): (b) rim-rim; (c) rim-disk; (d) disk-disk. }
	\label{fig:confcy}
\end{figure}

The first method for testing overlaps of hard cylinders was proposed by \citeauthor{Allen1993} \cite{Allen1993}, and also used by \citeauthor{Blaak1999} \cite{Blaak1999}. An alternative method was recently proposed by \citeauthor{Orellana2018} \cite{Orellana2018}. In the following, we use a refined version of this method outlined below.
The overlap of two cylinders can occur in either of the following three ways: disk-rim, rim-rim, and disk-disk (Figure~\ref{fig:confcy}).
Therefore, to ensure that the cylinders do not overlap,
we have to check whether the overlap occurs in one of those possible configurations, as detailed in Appendix \ref{sec:algorithm}.

\textcolor{black}{Preliminary simulations were initially performed to assess the computational effort of this algorithm compared with the hard spherocylinders counterpart. We found the present algorithm to be slightly slower - of the order of $20\%$ or less, and hence in line with Ref.~\onlinecite{Orellana2018} } 

We perform the less expensive test first, and progressively include additional more expensive ones. So we first check whether the two spheres (of diameter $L + D$) that encompass the cylinders overlap; if they do not, the cylinders cannot 
 overlap. If the encompassing spheres do overlap, the test is repeated for the spherocylinders enclosing the particles, using the standard algorithm 
 to calculate the shortest distance between two rods \cite{Vega1994}. Only if the spherocylinders overlap, the overlap between two cylinders is tested for. See Appendix \ref{sec:algorithm} for additional details.

\subsection{Order parameters}
To identify different thermodynamic phases, we rely on information based on global orientational and translational order, i.e., the nematic, smectic and \textcolor{black}{hexatic} order parameters,  on correlation functions such as the radial $g(\mathbf{r})$, parallel $g_{\parallel}(r_{\parallel})$ and perpendicular $g_{\perp}(r_{\perp})$ distribution functions, as well as on visual inspection of the simulation snapshots. 
\begin{figure*}[ht!]
  \centering
  \includegraphics[width = 1.0\textwidth]{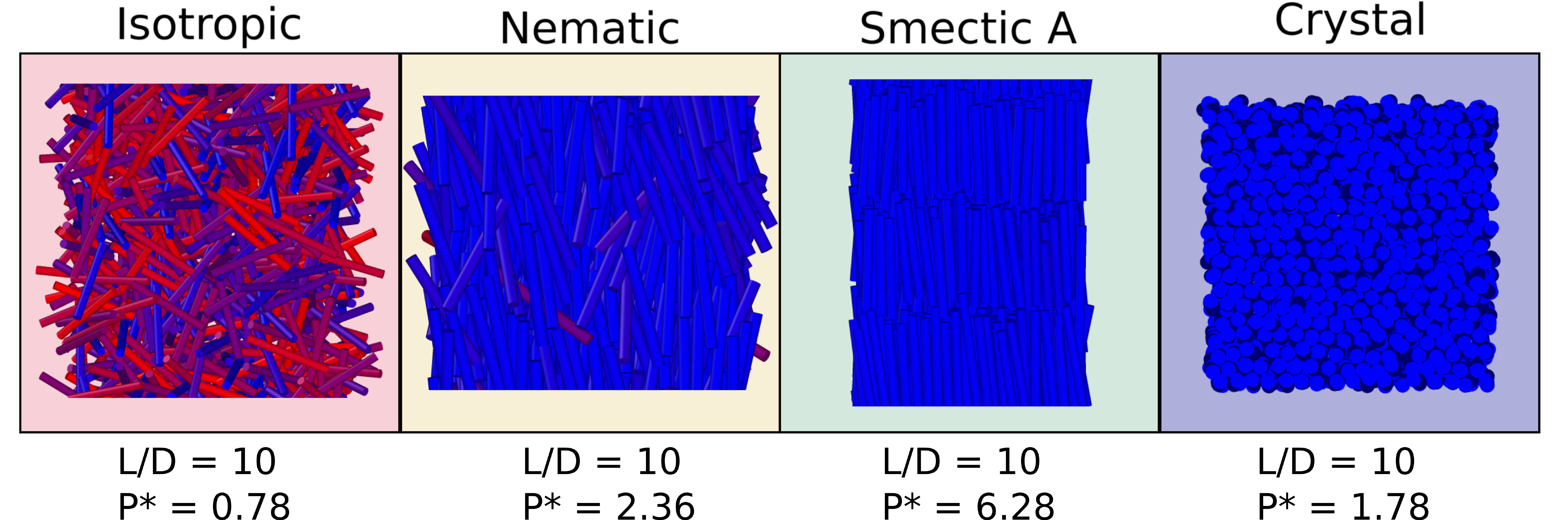}
  \caption{Representative snapshots of the thermodynamic phases found for HC for $L/D=10$: Isotropic (\textrm{I}), Nematic (\textrm{N}), Smectic A (\textrm{SmA}), and Crystal (\textrm{X}). Reduced corresponding pressures $P^{*}$ are displayed.}
  \label{fig:rodsphases}
\end{figure*}
Figure \ref{fig:rodsphases} displays representative snapshots of all different phases obtained in the  $L/D=10$ case -- all snapshots were obtained with the Ovito Software \cite{Stukowski2010} where different colours represent different orientations of the cylinders.
While the isotropic \textrm{I} phase is both positionally and orientationally disordered the nematic \textrm{N} phase is positionally disordered but orientationally ordered, and its presence can be inferred monitoring the nematic order parameter $P_2$. This is obtained as the largest eigenvalue of the tensor
\begin{eqnarray}
\label{sec2:eq1}
Q_{\alpha \beta} &=& \frac{1}{N}  \sum_{i=1}^N \frac{3}{2} \widehat{\mathbf{u}}_{\alpha}^{i} \widehat{\mathbf{u}}_{\beta}^{i} - \frac{1}{2} \delta_{\alpha \beta}
\end{eqnarray}
where $\alpha,\beta=x,y,z$.
The corresponding eigenvector then gives the main director $\widehat{\mathbf{n}}$.

In addition to the orientational order along one preferred direction $\widehat{\mathbf{n}}$, the smectic phase \textrm{SmA} is further characterised by a one-dimensional ordering (layering) along  $\widehat{\mathbf{n}}$ that is best captured by a combination of the radial distribution function
\begin{eqnarray}
\label{sec2:eq2}
g\left(r\right) &=& \frac{1}{N \rho} \frac{1}{4 \pi r^2} \left \langle \sum_{i=1}^N \sum_{j \ne i}^N \delta\left(r-r_{ij}\right) \right \rangle
\end{eqnarray}
as well as the parallel
\begin{eqnarray}
\label{sec2:eq3}
g_{\parallel} (r_{\parallel}) &=&
\frac{1} {N} \left \langle \frac{1}{ \rho L_x L_y}  
\sum_{i}^N \sum_{j \neq i}^N  \delta (r_{\parallel} - 
\mathbf{r}_{ij} \cdot \widehat{\mathbf{n}})  \right \rangle
\end{eqnarray}
positional correlation function. Here $\mathbf{r}_{i}$ is the center of mass of the $i$-th cylinder, and $\mathbf{r}_{ij}=\mathbf{r}_{j}-\mathbf{r}_{i}$, and $r_{ij}= \vert \mathbf{r}_{ij} \vert$. The smectic order parameter 
\begin{eqnarray}
\label{sec2:eq3b}
\left \langle \tau_1 \right \rangle &=& \left \vert \left \langle e^{\mathrm{i} 2\pi \frac{\mathbf{r} \cdot \widehat{\mathbf{n}}}{d}}\right \rangle \right \vert
\end{eqnarray}
also proves convenient. Here $\mathbf{r}$ is the position of a particle's centre of mass and $d$ the optimal layer spacing. Here and below, $\langle \ldots \rangle$ is the average over independent configurations at equilibrium. Then we have $\langle \tau_1 \rangle \approx 1$ in the smectic \textrm{SmA} phase and $\langle \tau_1 \rangle \approx 0$ elsewhere (phases with no layered structure).

By contrast, the columnar \textrm{C} phase is characterised by two-dimensional in-plane hexagonal order and one-dimensional positional disorder along $\widehat{\mathbf{n}}$. This is best captured by the perpendicular positional correlation function
\begin{eqnarray}
\label{sec2:eq4}
g_{\perp} (r_{\perp})&=&
\frac{1} {2 \pi r_{\perp} N} \left \langle \frac{1}{\rho L_z}  \sum_{i}^N \sum_{j \neq i}^N  \delta \left(r_{\perp} - 
\left \vert \mathbf{r}_{ij} \times \widehat{\mathbf{n}} \right \vert \right)  \right \rangle
\end{eqnarray}
positional correlation functions, as well as by the use of the hexatic (or bond) order parameter 
\begin{eqnarray}
\label{sec2:eq5}
\left \langle \psi_{6} \right \rangle &=& \left \langle \frac{1}{N} \sum_{j} \left \vert \frac{1}{n(j)} \sum_{\left \langle l m \right \rangle} e^{6 \mathrm{i} \theta_{lm}} \right \vert \right \rangle.
\end{eqnarray}
Here $\theta_{lm}$ is the angle that the  projection of the intermolecular vectors $\mathbf{r}_{jl}$
and $\mathbf{r}_{jm}$ onto the plane perpendicular to the director $\widehat{\mathbf{n}}$, 
$n(j)$ is the number of nearest-neighbours pairs of molecule within a single layer, and the sum
$\sum_{\langle lm \rangle}$ is over all possible pairs within the first coordination shell. With this definition, $\langle \psi_{6} \rangle \approx 1$ for hexagonal in-plane ordering and  $\langle \psi_{6} \rangle \approx 0$ otherwise. We refer to past literature (see e.g. \citeauthor{Kolli2016} \cite{Kolli2016}and references therein) for additional details.

Finally, the cubatic \textrm{Cub} phase corresponds to a long-range orientationally ordered phase without any positional order but with the presence of three equivalent perpendicular directions. In this phase, the particles form short stacks of typically few particles with neighbouring stacks tending to be perpendicular to one another along the three selected directions. While a suitable order parameter can be devised \cite{Duncan2009}, visual inspection is usually sufficient to unambiguously identify this phase. The details of the cubatic \textrm{Cub} phase will be discussed in Section \ref{sec:results}.

\begin{table}[h!]
  \centering
	\caption{Number of particles $N$ used in the simulations of rods.}
  \begin{tabular}{c c c c}
  \hline
      $L/D$ & $N$ & $L/D$ & $N$
      \\ \hline
      2.5 & 968
      &
      6.25 & 1350
      \\
      3.0 & 1152
      &
      6.5 & 1350
      \\
      3.25 & 1152
      &
      7.0 & 1536
      \\
      3.5 & 1352
      &
      7.5 & 1536
      \\
      5.0 & 1176
      &
      10.0 & 1944
      \\
      6.0 & 1350
      &
      \\ \hline
    \label{rodsnld}     
  \end{tabular}
\end{table}

\section{Results}
\label{sec:results}
\subsection{Cylindrical rods $L/D>1$}
\begin{figure}[ht!]
    \centering
    \subfigure[]{\includegraphics[width = 0.45\textwidth]{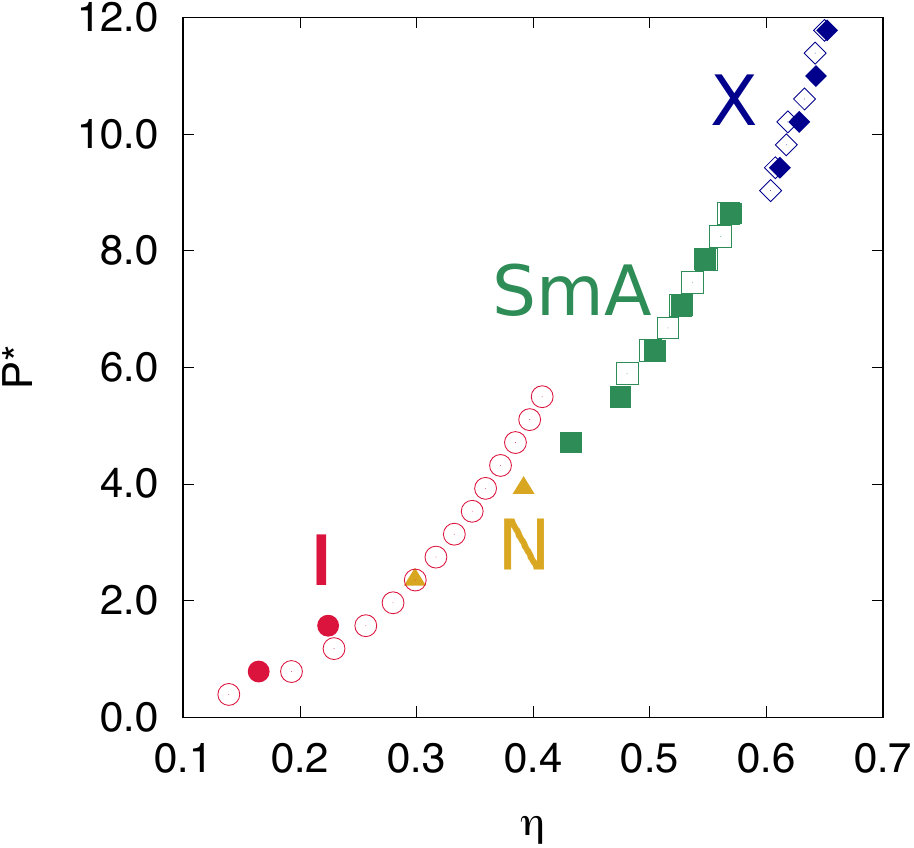}}  
    \subfigure[]{\includegraphics[width = 0.45\textwidth]{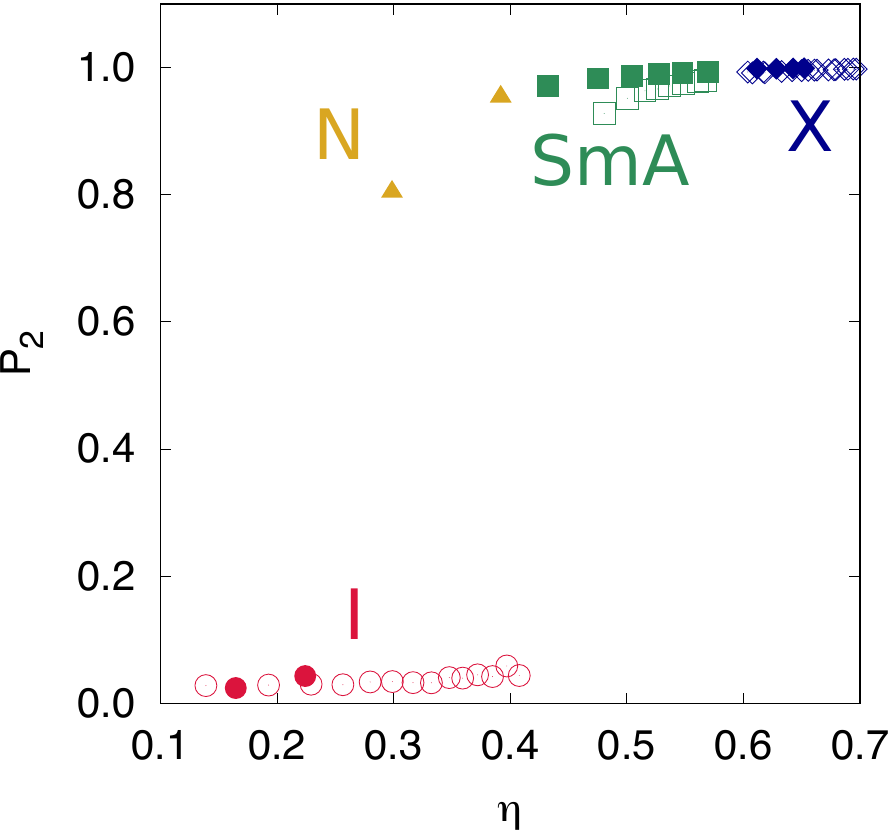}}   
     \caption{(a) Reduced pressure $P^{*}$ versus cylinder volume fraction $\eta$. Open symbols: $L/D = 5$, closed symbols: $L/D = 10$; (b) Nematic order parameter $P_2$ versus volume fraction $\eta$ for both $L/D=10$ and $L/D=5$. Same symbols as above. The different symbols and colours refer to different mesophases, as detailed in Fig~\ref{fig:rodsphases}.}
     \label{fig:EOS_cylinders}
    \end{figure}

We first consider the prolate case, i.e. cylindrical rods with $L/D>1$. Figure \ref{fig:EOS_cylinders} (a) depicts the reduced pressure $P^{*}$ as a function of the volume fraction $\eta$ (i.e., the equation of state) for $L/D=5$ and $L/D=10$. Figure \ref{fig:EOS_cylinders} (b) shows also the corresponding orientational order parameter $P_2$ again as a function of the volume fraction $\eta$.

In the case $L/D=5$ (open symbols) the system is in an isotropic phase \textrm{I} until $\eta \approx 0.4$, then switches to a smectic \textrm{SmA} phase, and then to a crystal \textrm{X}. The same sequence of phases is also found for the large aspect ratio $L/D=10$ (closed symbols) but with transitions shifted to lower $\eta$, and with the additional presence of a nematic \textrm{N} phase in the region $0.3 \le \eta \le 0.4$. The isotropic-nematic transition is signalled by an abrupt jump in the nematic order parameter $P_2$ and by a discontinuity in the equation of state as shown in Figure \ref{fig:EOS_cylinders} (a). 

Here it is worth to notice that our definition of crystal phase \textrm{X} includes the so-called smectic \textrm{SmB} phase, another name often used in this framework \cite{Kolli2016}, that is a smectic \textrm{SmA} phase with additional in-plane long-range hexagonal order \cite{Grelet2014}, thus hardly distinguishable from a crystal phase due to the finite size of the simulations box.

\begin{figure}[ht!]
    \centering
    \subfigure[]{\includegraphics[width = 0.45\textwidth]{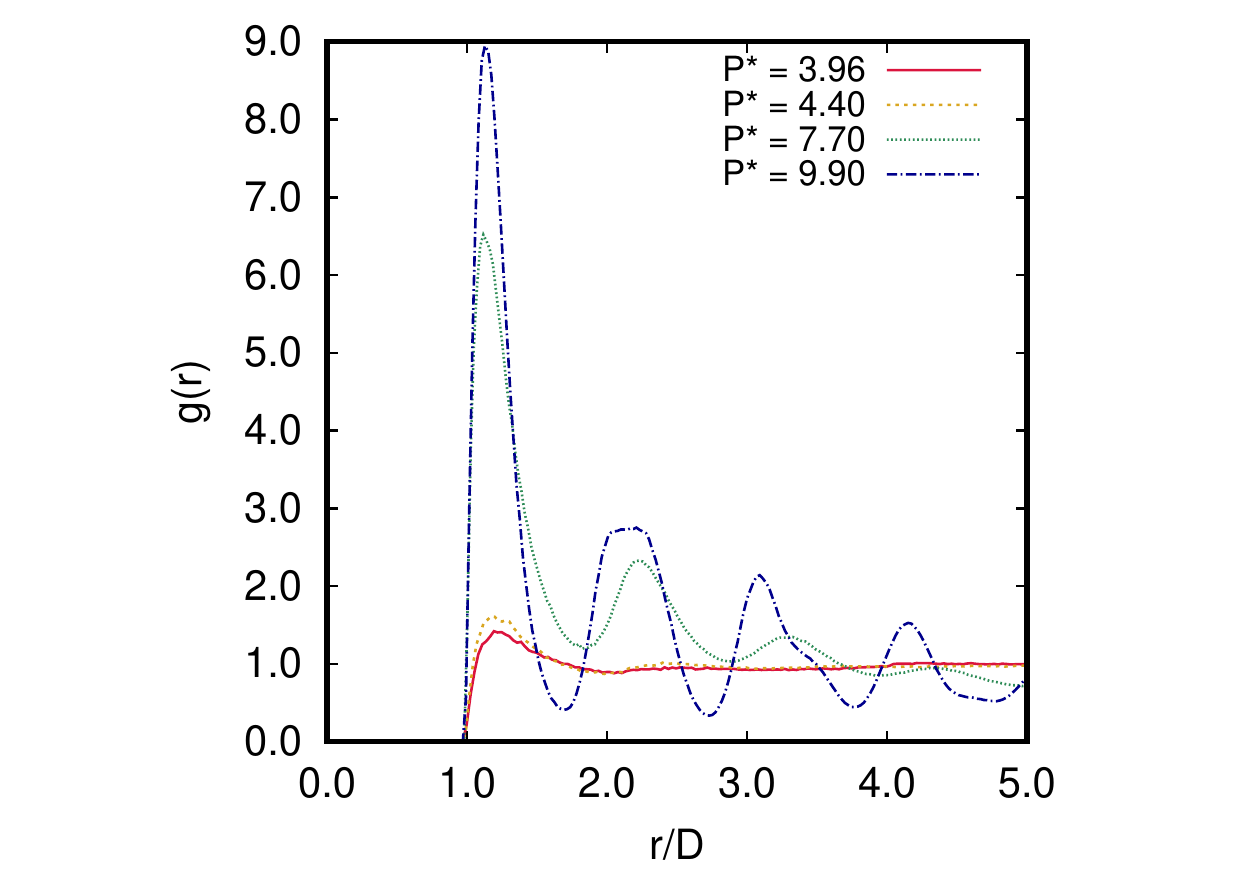}}
    \subfigure[]{\includegraphics[width = 0.45\textwidth]{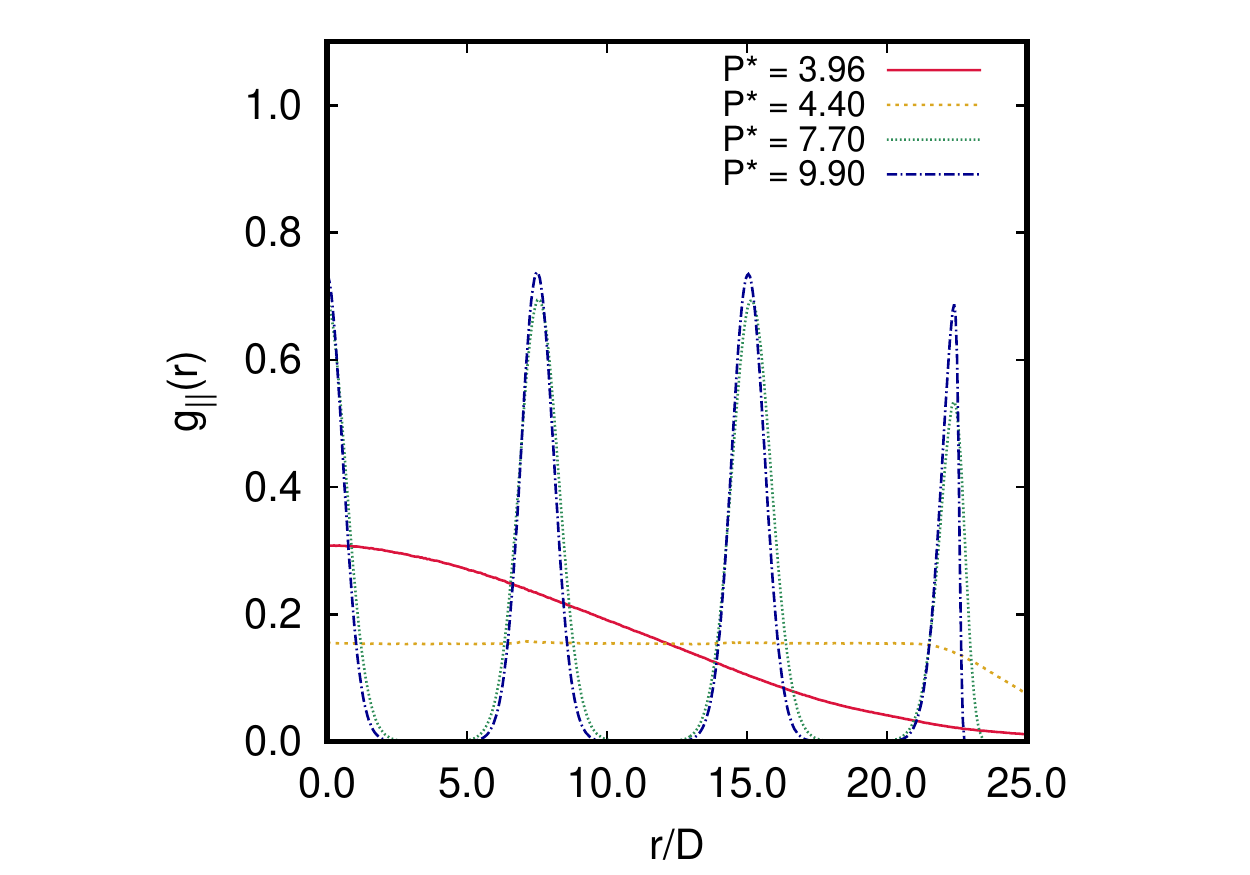}} 
    \subfigure[]{\includegraphics[width = 0.45\textwidth]{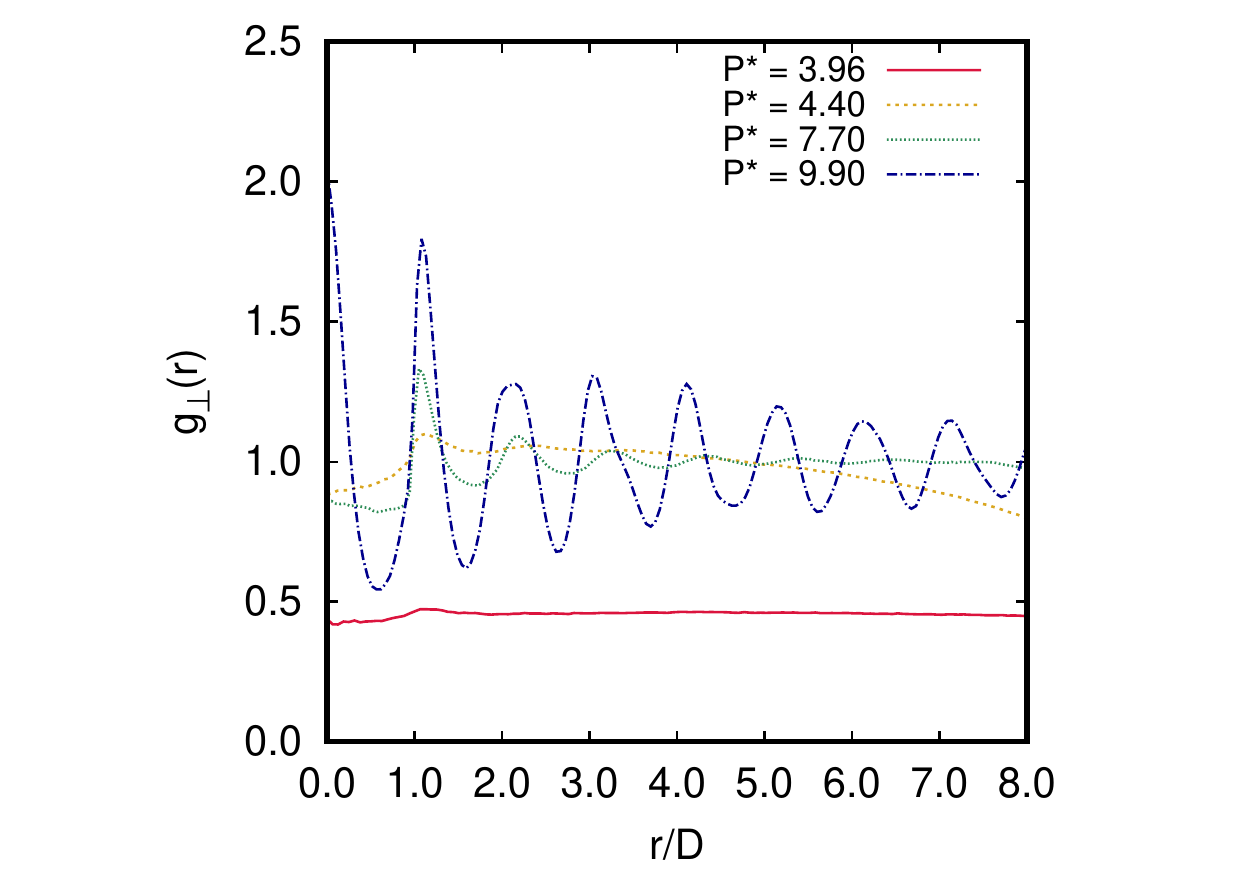}}
    \caption{
    Distribution functions of cylinders with $L/D = 7.0$. $P^* = 3.96$ continuous red line (\textrm{I}); $P^* = 4.40$ yellow dashed line (\textrm{N}); $P^* = 7.70$ green dotted line (\textrm{SmA}); $P^* = 9.90$ dash-dotted line (\textrm{X}). \textcolor{black}{Note that $r=\vert \mathbf{r}\vert$ in (a), $r=\vert \mathbf{r}_{\parallel}\vert$  in (b), and $r=\vert \mathbf{r}_{\perp}\vert$ in (c).}}
    \label{fig:l7df}
\end{figure}          

Additional insights can be obtained by looking at the correlation functions at an intermediate aspect ratio $L/D=7$ -- see Figure \ref{fig:EOS_cylinders_l7} for the analogue of Figure \ref{fig:EOS_cylinders} in the case $L/D=7$. Figure~\ref{fig:l7df} presents the corresponding radial $g(r)$ (a), parallel $g_{\parallel}(r_{\parallel})$ (b), and perpendicular $g_{\perp}(r_{\perp})$ (c) distribution functions of cylinders with $L/D = 7$ for increasing pressures. 

At $P^* = 3.96$ (continuous red line) all correlation functions are featureless, indicating the presence of an isotropic \textrm{I} phase. As pressure is increased up to $P^* = 4.40$ (yellow dashed line), the correlation functions do not show any significant change but the nematic order parameter $P_2$ (see Figure \ref{fig:EOS_cylinders_l7} (b)) shows an abrupt upswing, signalling the onset of a nematic \textrm{N} phase. 

At $P^* = 7.70$ (green dotted line), both the radial distribution function $g(r)$ and the parallel correlation function $g_{\parallel}(r_{\parallel})$ display a clear periodicity consistent with a smectic \textrm{SmA} ordering.
The absence of regular oscillations in the perpendicular correlation function $g_{\perp}(r_{\perp})$ confirms the radial liquid-like order of the mesophase, which therefore does not correspond to the crystal \textrm{X} phase. The latter phase is eventually reached at $P^* = 9.90$ (dash-dotted line) as shown by the characteristic periodicities for all directions in the  $g_{\perp}(r_{\perp})$ as well as in $g(r)$ and $g_{\parallel}(r_{\parallel})$.

It comes as no surprise that the low-$\eta$ behaviour of Hard Cylinders is qualitatively similar to the corresponding HSC counterpart \cite{Bolhuis1997}, with small quantitative differences for the smaller aspect ratio $L/D=5$. However, at high pressure and volume fraction, one possible important element of distinction between the two phase diagrams is the presence of a putative columnar phase that has already been demonstrated not to exist in the HSC counterpart \cite{Dussi18}. We explicitly addressed this problem following the method proposed by \citeauthor{Dussi18} \cite{Dussi18} who suggested that the apparent stabilisation of a columnar phase in HSCs could be ascribed to finite size effects when the number of layers is not sufficiently high compared to the aspect ratio $L/D$. For sake of consistency, we first reproduced the same results found in Ref. ~\onlinecite{Dussi18} for HSCs, and then applied the same method to the HC case. \textcolor{black}{We note that the metastability of the columnar phase for HSCs was also independently confirmed by \citeauthor{Liu2019} \cite{Liu2019} using a different method.}

\begin{figure}[ht!]
   \centering  
 \subfigure[~Unstable columnar phase]{\includegraphics[width = 0.5\textwidth]{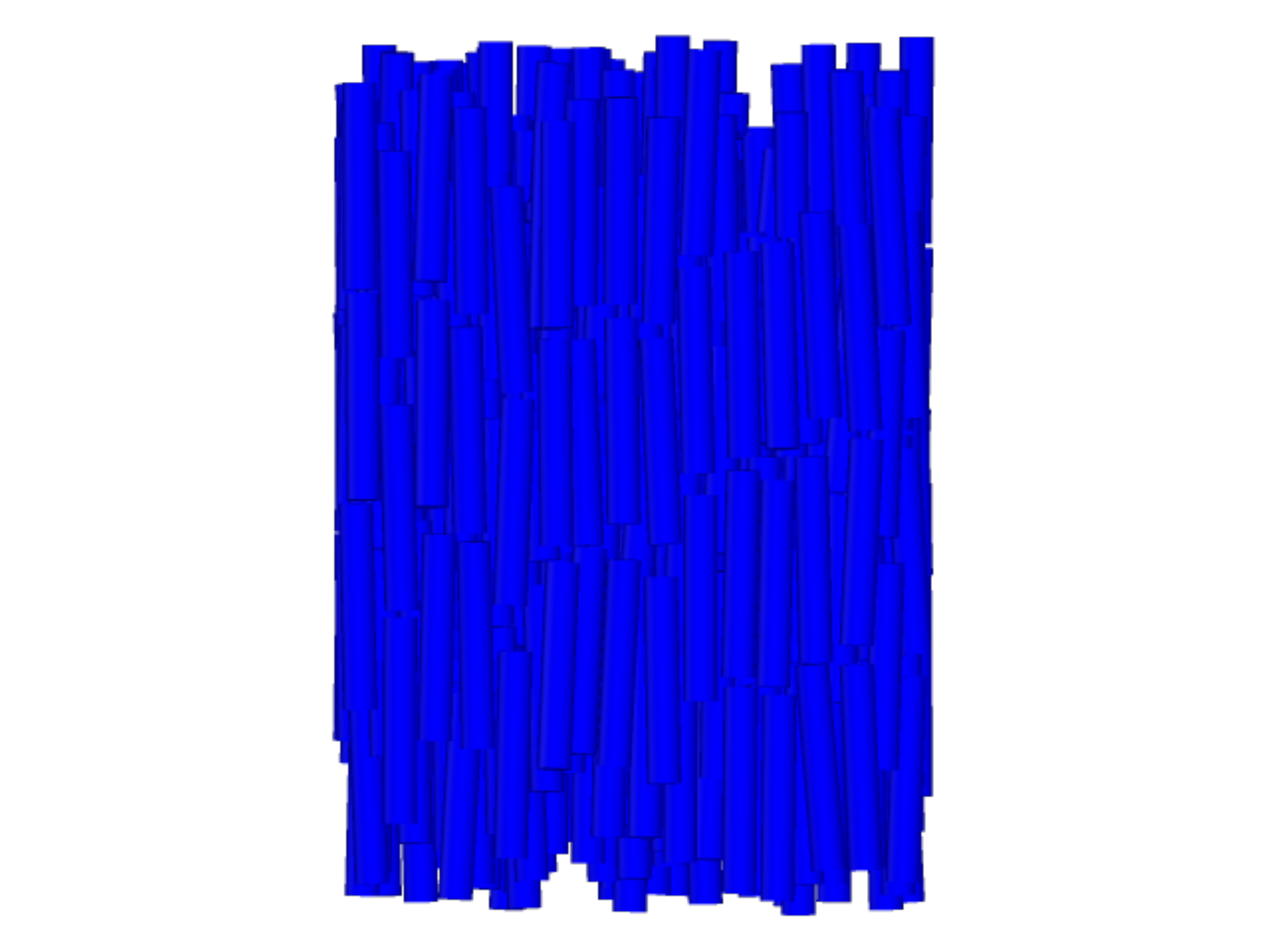}}   
 \subfigure[~Crystalline phase]{\includegraphics[width = 0.5\textwidth]{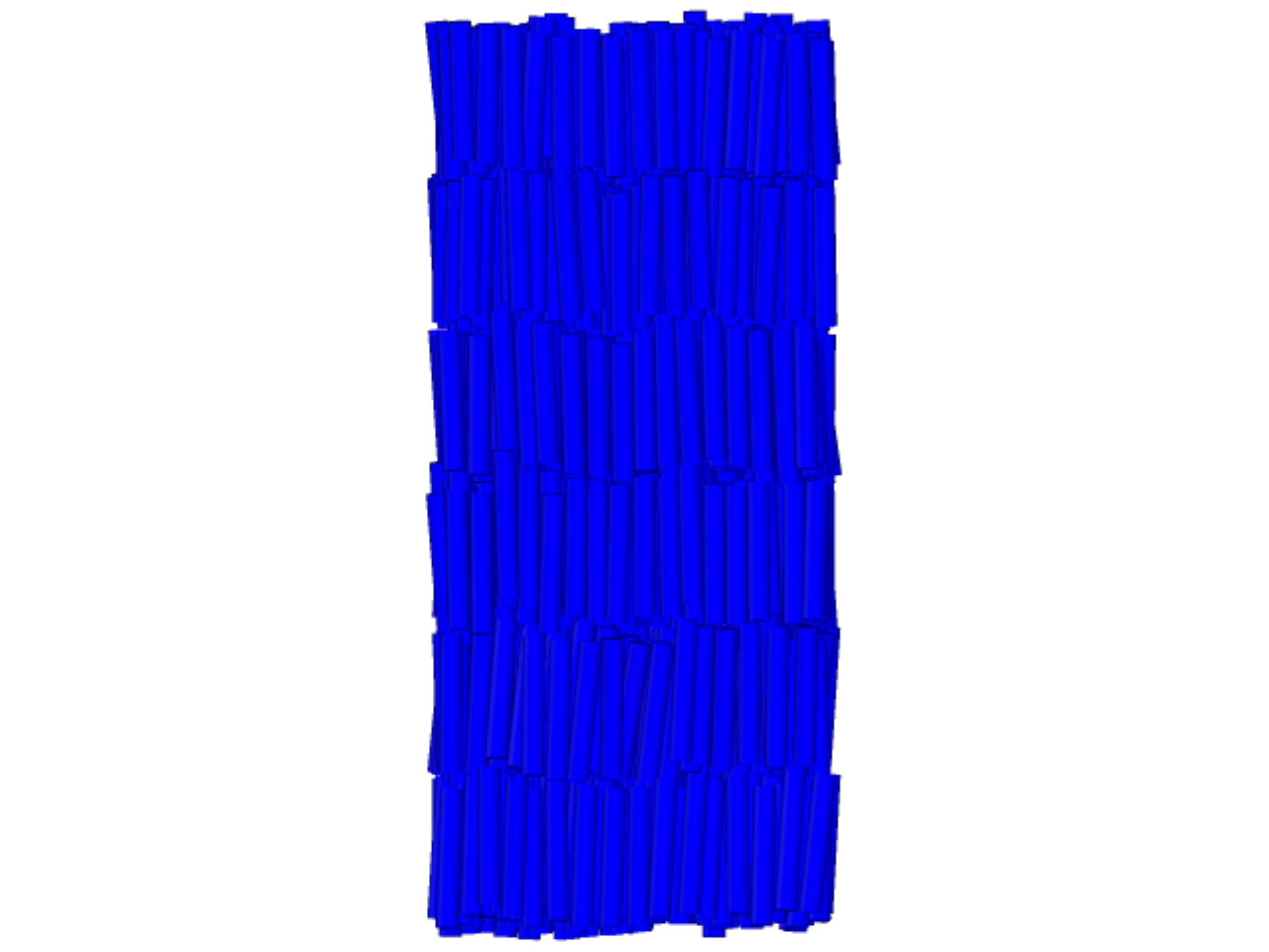}}                     
       \caption{Equilibrated configuration with aspect ratio $L/D = 6$ at packing fraction $\eta = 0.6$ with (a) $N=675$ initially distributed on two layers;(b) The same result with twice the particles exhibits a different structure.}
       \label{fig:falsecolumnar} 
    \end{figure}

The results obtained are presented in Figure \ref{fig:falsecolumnar}. Figure \ref{fig:falsecolumnar} (a) shows a production run  with aspect ratio $L/D = 6$ at packing fraction $\eta = 0.6$. In this case, both visual inspection and the behaviour of the corresponding correlation functions (see solid line in Figure \ref{fig:df} for results with $N=675$) strongly suggest the presence of a columnar phase.
However, if the number of particles is doubled along the director $\widehat{\mathbf{n}}$, the same calculation produces the final configuration shown in Figure \ref{fig:falsecolumnar} (b) that can clearly be classified as smectic \textrm{SmA} (see dotted line in Figure \ref{fig:df}  for results with $N=1350$). This shows that there is no stable columnar phase in HCs as in the \textcolor{black}{HSCs} case.  This effect is likely to be ascribed to the preference for finite size domains to arrange locally in columnar structures whose stability is eventually overwhelmed by long-range effects.


\begin{figure}[ht!]
  \centering
  \includegraphics[width = 0.5\textwidth]{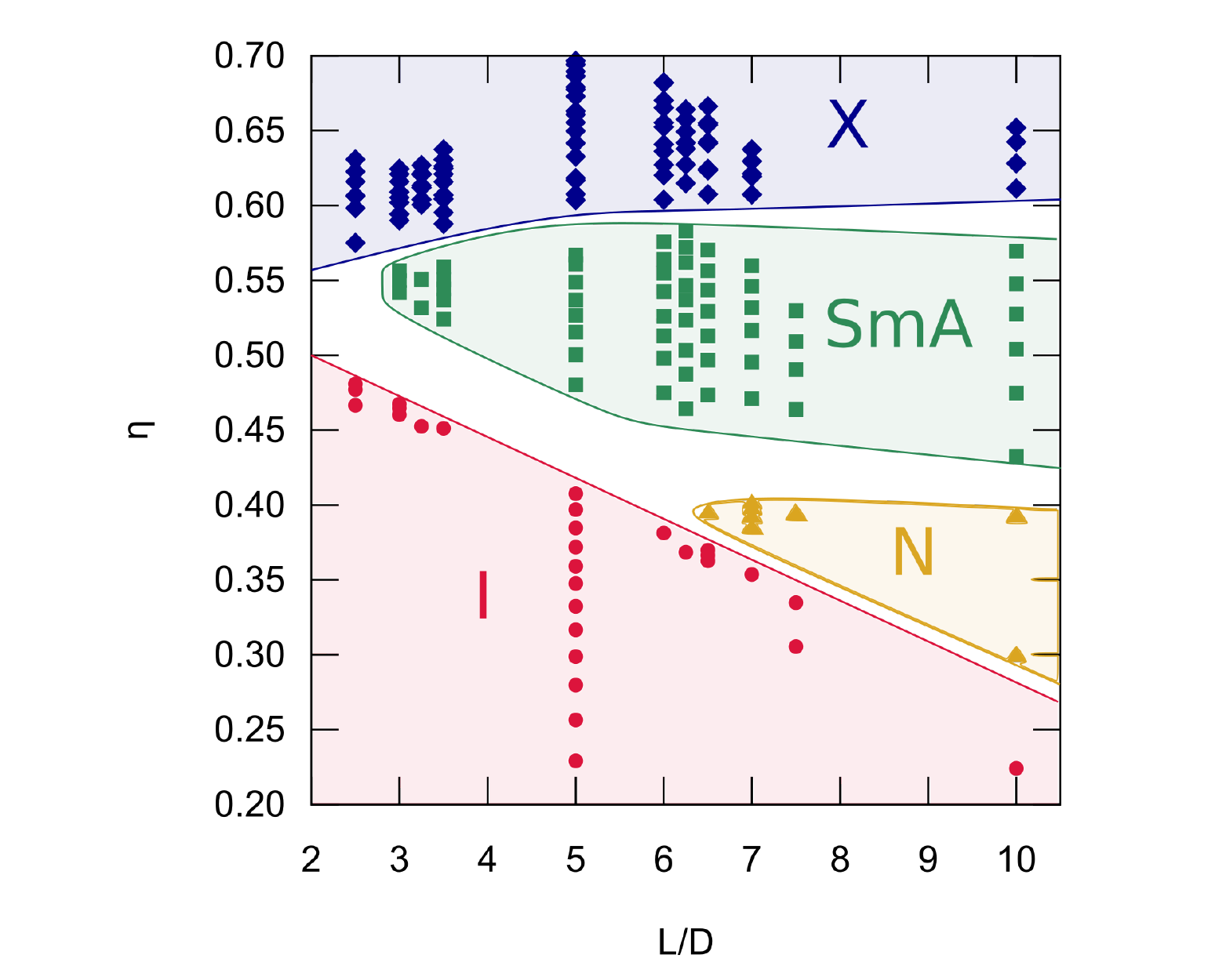}
  \caption{Computed phase diagram of hard cylinders of packing fraction $\eta$ versus aspect ratio $L/D$. Visible phases are isotropic (\textrm{I}), nematic (\textrm{N}), smectic (\textrm{SmA}), and crystal (\textrm{X}). Colour codes are as in Figure \ref{fig:rodsphases}.
  } 
  \label{fig:rodspd}
\end{figure}

A sketch of the final phase diagram for HCs in the plane packing fraction $\eta$ as a function of the aspect ratio ranging from $L/D= 2.5$ to $10$ is displayed in Figure~\ref{fig:rodspd}. The colour code used to represent different phases are outlined in
Figure~\ref{fig:rodsphases} that also presents representative snapshots 
of each phase. Here we employ the same classification as \citeauthor{Dussi18} \cite{Dussi18}.

Similar to hard spherocylinders, the system exhibits the isotropic (\textrm{I}), nematic (\textrm{N}), smectic A (\textrm{SmA}), and crystalline (\textrm{X}) phases. 
 Not surprisingly, this behaviour is similar to that of HSC \citep{Bolhuis1997,McGrother1996} but few differences are worth noticing. 
 
As in the case of HSC, no liquid crystal phases are observed below a critical aspect ratio $L/D \approx 3$.
 This fact can be easily rationalised via Onsager theory \cite{Onsager1949}, as the ratio between the covolume and volume of rods with lower $L/D$ are not sufficiently larger than that
 of a sphere, and the excluded volume effects then are 
 insufficient to promote an organised orientationally ordered phase.
 By contrast, at sufficiently high densities and aspect ratios,
 exclude volume effects tend to promote orientational order to increase the
 translational entropy, then minimising the free energy.
 
Accordingly, the system is in isotropic phase for any ratio $L/D$ below a certain packing fraction that decreases by increasing the rod aspect ratio, as shown in Figure \ref{fig:rodspd}. Upon increasing $\eta$, the first organised phase encountered is a smectic \textrm{SmA} phase in the range from $L/D = 3.25$  to $ L/D = 6$, and a nematic \textrm{N} phase above $L/D \approx 6$. This mirrors the HSC case where, however, the smectic \textrm{SmA} phase is limited to a very small range $3<L/D<4$. At higher packing fractions $\eta$, the system undergoes a smectic \textrm{SmA} to crystal \textrm{X} transition irrespective of the aspect ratio $L/D$.
 
 

The sketched phase diagram in Figure~\ref{fig:rodspd} prompts the existence of an isotropic-smectic-solid (\textrm{I-SmA-X}) triple point at $\eta \approx 0.55$ and \textcolor{black}{$L/D \approx 3.0$}, and an isotropic-nematic-smectic  (\textrm{I-N-SmA}) triple point at $\eta \approx 0.4$  and $L/D = 6.5$.

Interestingly, the location of the \textrm{I-N-SmA} is found at $L/D \approx 6.5$ and shifted to higher aspect ratios compared to that of HSCs that is found at $L/D \approx 3.7$ \cite{Bolhuis1997}. As a result, the nematic \textrm{N} phase stabilises at shorter aspect ratios in the HSC system when compared to its HC counterpart.

It is also interesting to notice that our results are compatible with both the \textrm{I}-\textrm{N} and the \textrm{N}-\textrm{SmA} being first-order transitions, thus mirroring what is known for HSC from the work by \citeauthor{Polson1997} \cite{Polson1997}. It would be interesting to pursue the same analysis carried out by these authors in the present case as well. The same consideration holds true for interesting analysis of the $L/D \to \infty$ Onsager limit that has been performed for the HSC case \cite{Bolhuis1997} that could be also replicated in this case.

\textcolor{black}{As a final remark we note that the length $L$ for HSCs corresponds to $L+D$ in case of HCs. This is important when comparing the corresponding phase diagrams and indeed it rationalizes why the isotropic-smectic \textrm{I-SmA} transition for HCs occurs at $L/D \approx 3$ whereas for HSCs occurs at $L/D \approx 4$. However, the tendency of a flat edge to promote the onset of a smectic \textrm{SmA} phase appears to be a general feature as also suggested by a recent study \cite{Marechal2017} on hard equilateral triangular prisms, where the particles feature also flat sides and the smectic \textrm{SmA} phase is shifted to considerably lower packing fractions as compared to HSCs.}

\subsection{Cylindrical disks $L/D<1$}
\begin{figure*}[ht!]
  \centering
  \includegraphics[width = 0.9\textwidth]{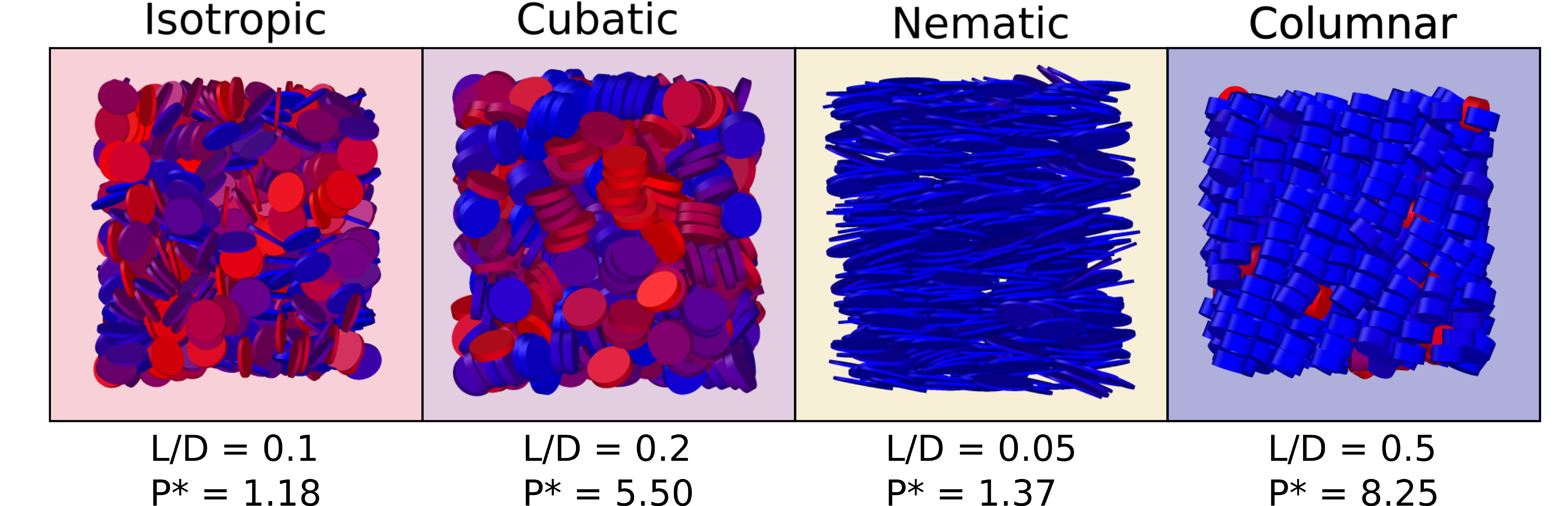}
  \caption{Representative snapshots of the different phases found in the oblate $L/D<1$ case: Isotropic (\textrm{I}), Cubatic (\textrm{Cub}), Nematic (\textrm{N}), Columnar (\textrm{C}). The corresponding values of aspect ratios $L/D$ and reduced pressure $P^{*}$ are also reported. }
    \label{fig:diskphases}
\end{figure*}

We now tackle the oblate case of cylindrical disks with $L/D<1$. One important advantage of dealing with cylinders is that this limit can be achieved with no solution of continuity, unlike the spherocylinders counterpart where this is not possible \cite{Bolhuis1997}. Figure~\ref{fig:diskphases} depicts the \textcolor{black}{four} different phases that we find in this case: a disordered isotropic \textrm{I}, a cubatic \textrm{Cub}, a nematic \textrm{N}, and a columnar \textrm{C}  phases, as detailed in Table \ref{dcolorcode} and illustrated in Figure \ref{fig:diskphases}. 
 \begin{table}[h!]
  \centering
  \caption{Colours and symbols used to represent disk phases.}
  	\label{dcolorcode}
  \begin{tabular}{c c c c}
      \hline
      Colour & Phase & Notation&Symbol
      \\ \hline
      red & isotropic & I & circle
      \\
      yellow & nematic & N & triangle
      \\
      purple & cubatic & Cub & squares
      \\
      blue & columnar & C & diamond
      \\ \hline
	\end{tabular}
\end{table}

\begin{figure}[ht!]
  \centering
  \includegraphics[width = 0.5\textwidth]{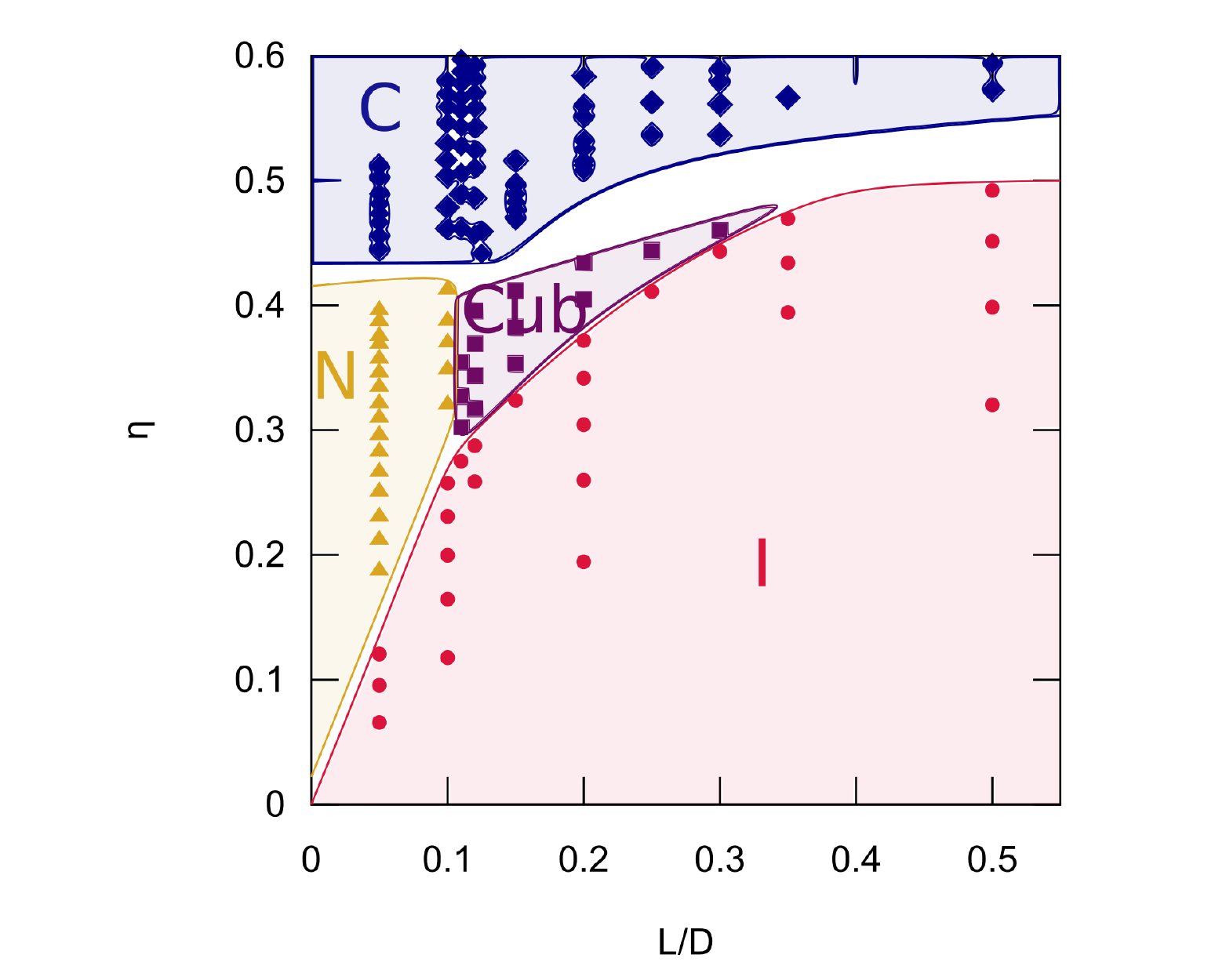}
  \caption{Phase diagram of the oblate hard cylindrical disk case $L/D<1$ in the volume fraction $\eta$ aspect ratio $L/D$ plane. Different phases are colour coded as detailed in Figure  \ref{fig:diskphases} and Table \ref{dcolorcode}.} 
  \label{fig:diskspd}
\end{figure}

\begin{table}[h!]
  \centering
\caption{Number of particles $N$ used in the simulations of disks.}
\label{disksnld}
  \begin{tabular}{c c c c}
	\hline
	$L/D$ & $N$ & $L/D$ & $N$
	\\ \hline
	0.05 & 540 
	&
	0.2 & 625
	\\ 
	0.1 & 640
	&
	0.25 & 864
	\\ 
	0.11 & 576
	&
	0.3 & 720
	\\ 
	0.12 & 528
	&
	0.35 & 612
	\\
	0.125 & 528
	&
	0.5 & 686
	\\
	0.15 & 825
    & &
	\\ \hline
  \end{tabular}
\end{table}
As in this case of prolate cylinders, we performed the same detailed analysis of the different obtained phases in terms of correlation functions and order parameters to derive the equation of states. Supplementary Figure S3 reports the reduced pressure $P^{*}$ and the $P_2$ nematic order parameter as a function of the packing fraction $\eta$ for both $L/D=0.2$ and $L/D=0.05$ as representative examples, from which one can obtain the corresponding phase diagram of Figure~\ref{fig:diskspd} in the volume fraction $\eta$ aspect ratio $L/D$ plane, that can be contrasted with its prolate counterpart of Figure \ref{fig:rodspd}. Here a range from $L/D= 0.05$ to $L/D=0.5$ has been analysed and in Figure \ref{fig:diskphases} representative snapshots of different phases are depicted colour-coded according to Table~\ref{dcolorcode}, in analogy with the discussion of the prolate case $L/D>1$.
 
For aspect ratios $0.3<L/D<0.5$ there is a direct transition from an isotropic \textrm{I}  to a columnar \textrm{C} phase upon increasing $\eta$ above $\approx 0.4$. In the columnar phase the disks are arranged on a hexagonal lattice in the direction perpendicular to the main director $\widehat{\mathbf{n}}$ but their centres of mass are disorderly distributed in space.
For smaller aspect ratios  $0.1<L/D<0.3$, a cubatic \textrm{Cub} phase appears between the \textrm{I} and \textrm{C} phase. In the cubatic phase, the disks tend to assemble in short stacks of about four or five units, with neighbouring columns perpendicular to each other. This differs from the cubic phase because it lacks translational order \cite{Veerman1992}. At even smaller aspect ratio ($L/D\le 0.1$) the cubatic \textrm{Cub} phase is replaced by a nematic \textrm{N} phase up to $\eta \approx 0.4$ and by a columnar phase at higher $\eta$.
At even higher packing fractions $\eta$ we did observe the formation of a crystal phase \textrm{X} but the location of corresponding boundaries would require a specific investigation that was not pursued in the present paper. 

\begin{figure}[ht!]
  \centering
  \subfigure[]{\includegraphics[width = 0.45\textwidth]{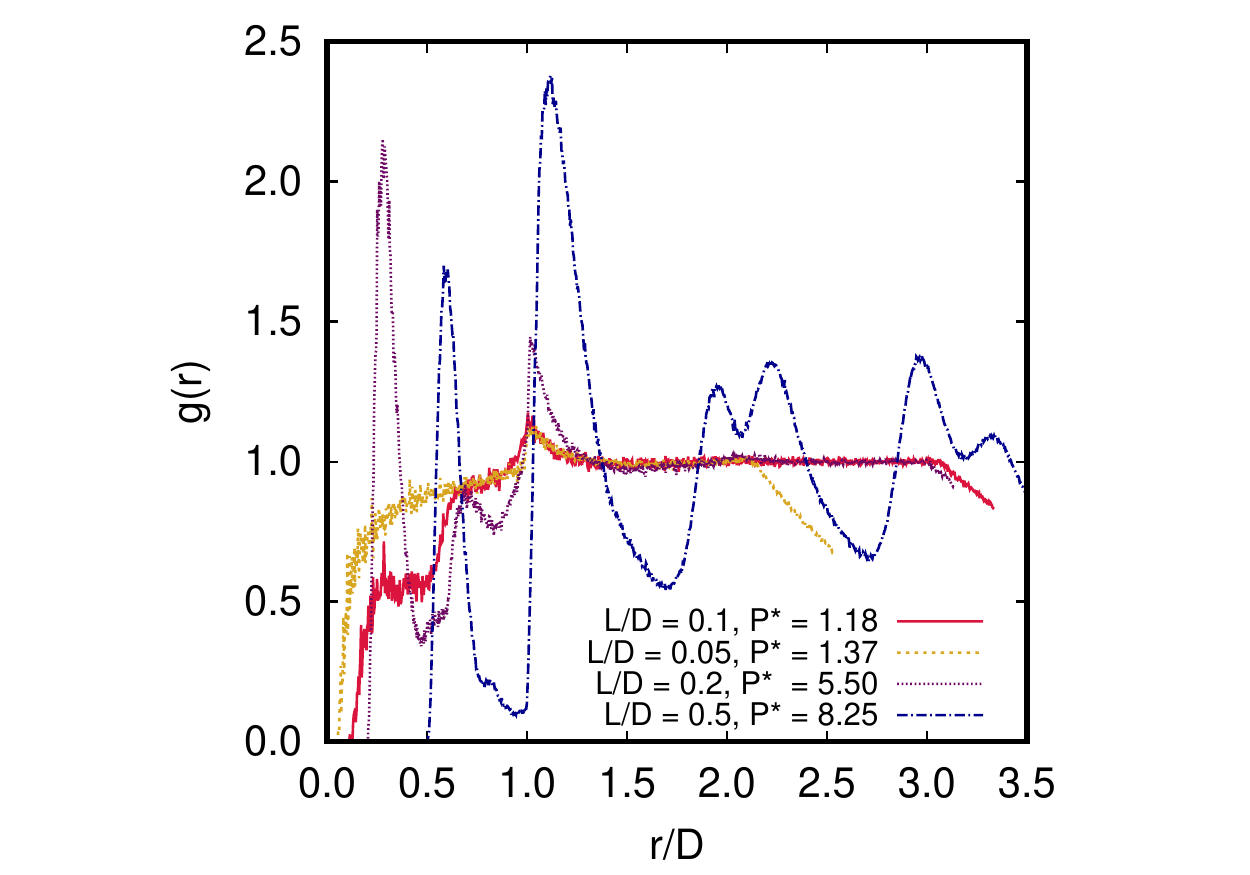}}
  \subfigure[]{\includegraphics[width = 0.45\textwidth]{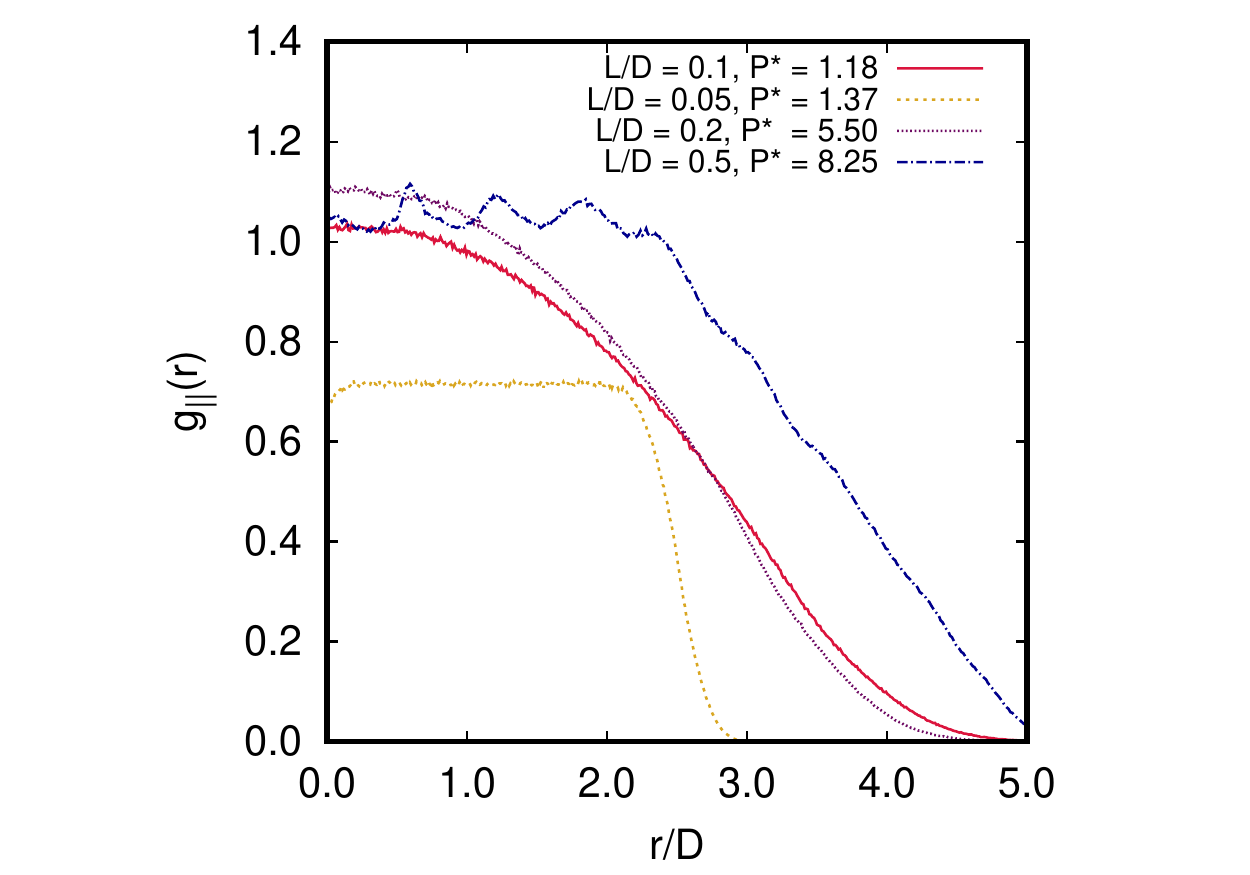}}
  \subfigure[]{\includegraphics[width = 0.45\textwidth]{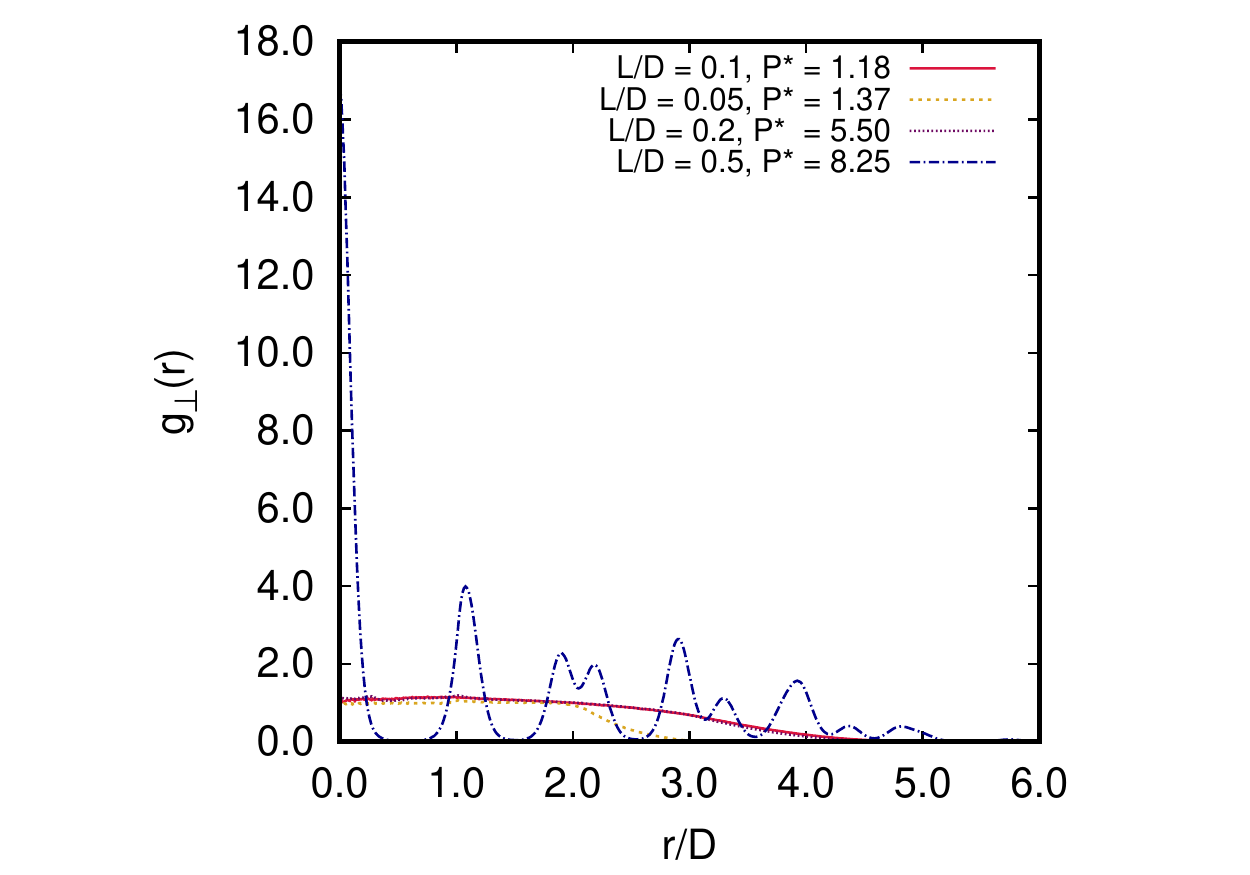}}
  \caption{Distribution functions of hard cylindrical disks.
  $L/D = 0.1$ and $P^* = 1.18$ red continuous line \textrm{I}; $L/D = 0.05$ and $P^* = 1.37$ blue dashed line \textrm{C}; $L/D = 0.2$ and $P^* = 5.50$  dotted purple line \textrm{Cu}; $L/D = 0.5$ and $P^* = 8.25$
	dash-dotted yellow line \textrm{N}. Colour code is outlined in Table~\ref{dcolorcode}.\textcolor{black}{Here again $r=\vert \mathbf{r}\vert$ in (a), $r=\vert \mathbf{r}_{\parallel}\vert$  in (b), and $r=\vert \mathbf{r}_{\perp}\vert$ in (c).}}
	\label{fig:disksdf}
  \end{figure}

All these transitions can be best inferred by looking at the correlation functions as shown in Figure \ref{fig:disksdf} (see Fig. \ref{fig:diskphases} for the corresponding snapshots). In the \textrm{I} phase (red continuous line) the radial distribution function $g(r)$ displays a flat behaviour for $r>D$, indicating the absence of short-range aggregation, a feature confirmed by the behaviour of both $g_{\parallel}(r_{\parallel})$ and $g_{\perp}(r_{\perp})$. By contrast, the columnar \textrm{C} phase (blue dashed line) displays characteristic regular oscillations in $g(r)$ and $g_{\perp}(r_{\perp})$, but the behaviour of $g_{\parallel}(r_{\parallel})$ is irregular, indicating the absence of a one-dimensional ordering along the main director $\widehat{\textbf{n}}$.
Likewise, the radial distribution function $g(r)$ of the cubatic phase (dotted purple line) is quite different from both the \textrm{I} and \textrm{C} phases, \textcolor{black}{while the nematic order parameter $P_2$ is close to zero for  \textrm{I}, and \textrm{Cub}. }
An evidence of the formation of short stacks is the higher peak at short distances ($L/D < r/D < 2L/D$) in the radial distribution function $g(r)$ of the cubatic \textrm{Cub} phase (purple line in Figure~\ref{fig:disksdf}), which is significantly smaller in the the $g(r)$ of an isotropic phase (red line in Figure~\ref{fig:disksdf}).
Finally, the onset of the nematic \textrm{N} phase (dash-dotted yellow line) is signalled by the significant oscillation of the radial distribution function $g(r)$ and by the abrupt upswing in the nematic order parameter $P_2$, as shown in supplementary materials.

At variance with the prolate $L/D>1$ counterpart, in the oblate case three triple points appear. The \textrm{I-N-Cub} triple point occurs at $L/D \approx 0.1$ and $\eta \approx 0.3$. The \textrm{N-Cub-C} triple point is approximately located at $L/D \approx 0.1$ and $\eta \approx 0.4$. Finally, the \textrm{I-Cub-C} triple point has approximate coordinates $L/D \approx 0.35$ and $\eta \approx 0.45$.

Our results qualitatively agree with density function calculations by \citeauthor{Wensink2009}\cite{Wensink2009} who predicted the existence of a nematic region for flat disks that becomes progressively narrower as $L/D$ increases. The same authors also predicted a transition from the isotropic phase directly to the columnar \textrm{C} phase, in agreement with our results. At a more quantitative level the predicted volume fractions $\eta_{IN} \approx \pi L/D$ for the isotropic-nematic transition and $\eta_{NC} \approx 0.4$ for the nematic-columnar \textrm{N-C}, as somewhat larger than those found in the present study.

At variance of these theoretical findings, our results indicate also the existence of a cubatic \textrm{Cub} phase, in agreement with the results by \citeauthor{Veerman1992}\cite{Veerman1992}, as well as by \citeauthor{Duncan2009} \cite{Duncan2009}, in simulations of cut spheres, and by \citeauthor{Blaak1999} \cite{Blaak1999} in simulations of hard cylinders. Where direct comparison with the above two papers is possible, we find complete agreement between their results and ours. 
 
\citeauthor{Duncan2009}\cite{Duncan2009} simulated cut spheres of
$L/D = 0.1$, $0.15$, $0.2$, $0.25$ and $0.3$ and, despite the differences 
in shape, our results are very similar to theirs. These authors showed that 
there is a nematic \textrm{N} but no cubatic \textrm{Cub} phase at $L/D = 0.1$, and that the opposite 
is true for $L/D \ge 0.15$. Figure~\ref{fig:diskspd} shows that for HCs a cubatic \textrm{Cub} phase is already present at $L/D = 0.11$, as the nematic \textrm{N} phase vanishes. The cubatic \textrm{Cub} phase is present until $L/D \approx 0.3 $ whereas only the isotropic \textrm{I} and columnar \textrm{C} phases exist at larger aspect ratios.

As in our case, \citeauthor{Blaak1999}\cite{Blaak1999} investigated a system of HC with $L/D = 0.9$ and did not find any cubatic phase. Our results explain this finding by showing that $L/D = 0.9$ is a too large aspect ratio to support a cubatic phase that is however present at smaller aspect ratio  $0.1<L/D<0.3$, as shown in Figure \ref{fig:diskspd}.

  \section{Conclusions}
  \label{sec:conclusions}
In this paper we have used isobaric $(NPT)$ Monte Carlo simulations to study the phase diagram of a system of $N$ hard cylinders as a function of their aspect ratio and volume fraction. To achieve this, we have implemented a new and efficient overlap test for hard cylinders that compares well with those existing in the literature \cite{Blaak1999,Orellana2018}. This allows us to study the complete phase diagrams in the aspect ratio versus volume fraction for both the prolate $L/D>1$ and the oblate $L/D<1$ cases.

In the prolate case $L/D>1$, we find a phase diagram very similar to the hard spherocylinders counterpart, featuring the presence of a nematic \textrm{N} and a smectic \textrm{SmA} phases, in addition to the isotropic \textrm{I} and the crystal \textrm{X} phases, as well as two \textrm{I-Sm-X} and \textrm{I-N-SmA} triple points.
As in the spherocylinder case \cite{Dussi18}, we have shown that the appearance of a columnar \textrm{C} phase can be traced back to a finite-size effect and that it disappears for sufficiently large systems. Our simulations confirm the lack of existence of a stable columnar \textrm{C} phase, that was nevertheless predicted by density functional theory \cite{Wensink2009}.

In the oblate case $L/D<1$, we identified the presence of a columnar \textrm{C}, a nematic \textrm{N} and a cubatic \textrm{Cub} phase, in agreement with theoretical prediction \cite{Wensink2009}, as well as with past numerical simulations of cut spheres \cite{Veerman1992,Duncan2009} and of hard cylinders \cite{Blaak1999}. In the latter case, we have provided an explanation of the failure of past simulations of identifying the cubatic \textrm{Cub} phase that can be ascribed to the too large aspect ratio used in these simulations. Interestingly, the phase diagram also includes three \textrm{I-N-Cub}, \textrm{N-Cub-C}, and \textrm{I-Cub-C} triple points.

The present study paves the way to tackling more complex systems building upon cylindrical shapes that are of experimental interest, such as hard cylinders interacting via a Yukawa tail \cite{Grelet2014}, as well as hard cylinders with short-range directional attractions \cite{Livolant2012a,Repula2019}. Such investigations are underway and will be reported elsewhere.


\acknowledgments{We are indebted to Cristiano \textcolor{black}{De} Michele and Maria Barbi for useful discussions. The use of the SCSCF multiprocessor cluster at the Universit\`{a} Ca' Foscari Venezia and of the LESC cluster at the University of Campinas are gratefully acknowledged. The work was supported by MIUR PRIN-COFIN2017 \textit{Soft Adaptive Networks} grant 2017Z55KCW  and Galileo Project 2018-39566PG (AG), by S\~ao Paulo Research Foundation (FAPESP) grant no. 2018/02713-8, and
by the Coordena\c{c}\~ao de Aperfei\c{c}oamento de Pessoal de N\'ivel Superior - Brasil (CAPES) - Finance Code 001. \textcolor{black}{FR acknowledges IdEx Bordeaux (France) for financial support.} JL gratefully acknowledges the hospitality of the Ca' Foscari University of Venice where part of this work was carried out. \textcolor{black}{The authors would like to acknowledge the contribution of the Eutopia COST Action CA17139.}}

\section*{Data Availability}
The data that supports the findings of this study are available within the article and its supplementary material.

%
%

%


\appendix
\section{Algorithm to check overlap between two cylinders}
\label{sec:algorithm}

\subsubsection{Parallel Cylinders}
If two cylinders are parallel, the overlap can occur between disk-disk or rim-rim only, and it can be easily checked. For each particle pair we define four vectors starting from the the vector joining the two centres of mass, $\mathbf{r}_{12}$. We do this by extracting its parallel and perpendicular component with respect to the director $\mathbf{\widehat{u}}_j$ of each particle, i.e., 
	\begin{equation}
		\begin{aligned}
			\mathbf{r}_{1\parallel} = &~(\mathbf{r}_{ij}\cdot \mathbf{\widehat{u}}_1)\mathbf{\widehat{u}}_1\\ 
			\mathbf{r}_{2\parallel} = &~(\mathbf{r}_{ij}\cdot \mathbf{\widehat{u}}_2)\mathbf{\widehat{u}}_2\\ 
			\mathbf{r}_{1\bot} = &~\mathbf{r}_{12} - (\mathbf{r}_{12}\cdot \mathbf{\widehat{u}}_1)\mathbf{\widehat{u}}_1\\
			\mathbf{r}_{2\bot} = &~\mathbf{r}_{12} - (\mathbf{r}_{12}\cdot \mathbf{\widehat{u}}_2)\mathbf{\widehat{u}}_2
		\end{aligned}
		\label{decvec}
	\end{equation}
	
In a parallel configuration, the directors can either be the same of one the opposite of the other. The overlap occurs if \textit{all} the following conditions are satisfied:
\begin{equation}
	\begin{aligned}
		|\mathbf{r}_{1 \parallel}| & \leq L \\
		|\mathbf{r}_{2 \parallel}| & \leq L \\
		|\mathbf{r}_{1 \bot}| & \leq D \\
		|\mathbf{r}_{2 \bot}| & \leq D
	\end{aligned}
	\label{parovp}
\end{equation}

When the cylinders are \textit{exactly} parallel, $\mathbf{\widehat{u}}_1 = \pm \mathbf{\widehat{u}}_2$, and half of the conditions above are redundant since $|\mathbf{r}_{1 \bot}| = |\mathbf{r}_{2 \bot}|$ and $|\mathbf{r}_{1 \parallel}|=|\mathbf{r}_{2 \parallel}|$; when implementing computer code, however, one has to include tolerances and care must be taken in in handling these conditions consistently.

\subsubsection{Rim-rim overlap}

Since the overlap between spherocylinders is the first test that is
done, and the rim of a spherocylinder is similar to the rim of a 
cylinder, if the rims of two spherocylinders do overlap, than the two cylinders 
will certainly overlap as well. Hence, having performed the spherocylinder overlap test, we now check if the overlap occurs in a rim-rim configuration.

To that end, we define the vectors 
$\mathbf{V}_1 = -\mathbf{r}_{12} + \lambda \mathbf{\widehat{u}_1}$ and 
$\mathbf{V}_2 = \mathbf{r}_{12} + \mu \mathbf{\widehat{u}_2}$, where the numbers 
$\lambda$ and $\mu$, consistently with Ref.~\onlinecite{Vega1994}, identify the points of closest approach between the axes of the two cylinders. These values are calculated using the \citeauthor{Vega1994}'s algorithm\cite{Vega1994}, which we implement in the spherocylinder overlap test.
If the cylinders are in a rim-rim configuration, the two conditions below must both be satisfied:
\begin{equation}
\begin{aligned}
|\mathbf{V}_1\cdot \mathbf{\widehat{u}_2}|< & L/2\\
|\mathbf{V}_2\cdot \mathbf{\widehat{u}_1}|< & L/2
\end{aligned}
\end{equation}

\begin{figure}[htbp]
  \centering
  \subfigure[Rim-rim configuration]{\includegraphics[width = 0.2\textwidth]{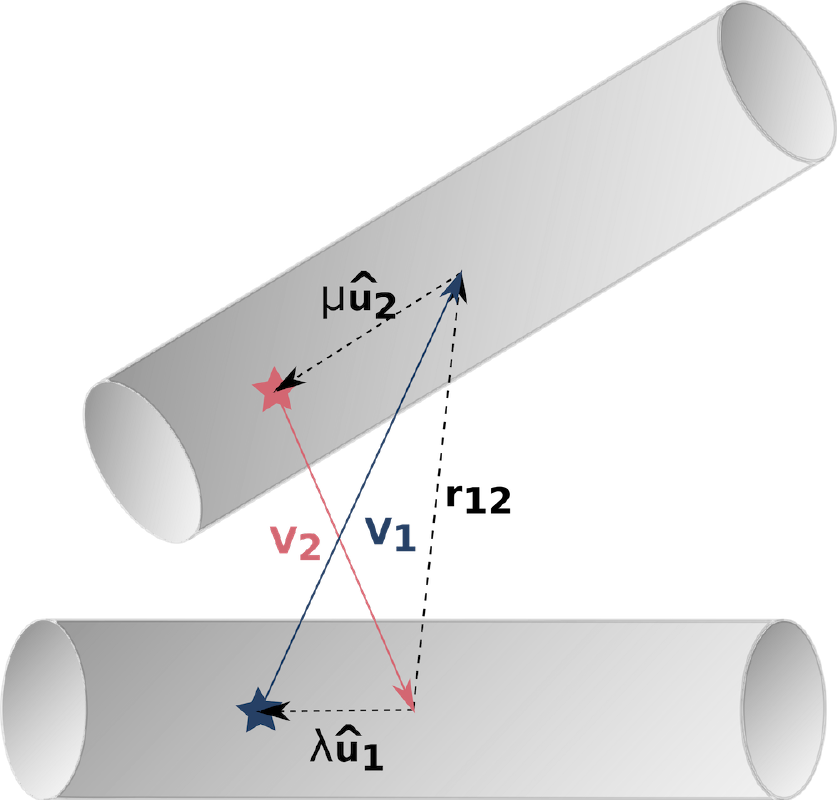}}
  \centering
  \subfigure[Disk-rim configuration]{\includegraphics[width = 0.2\textwidth]{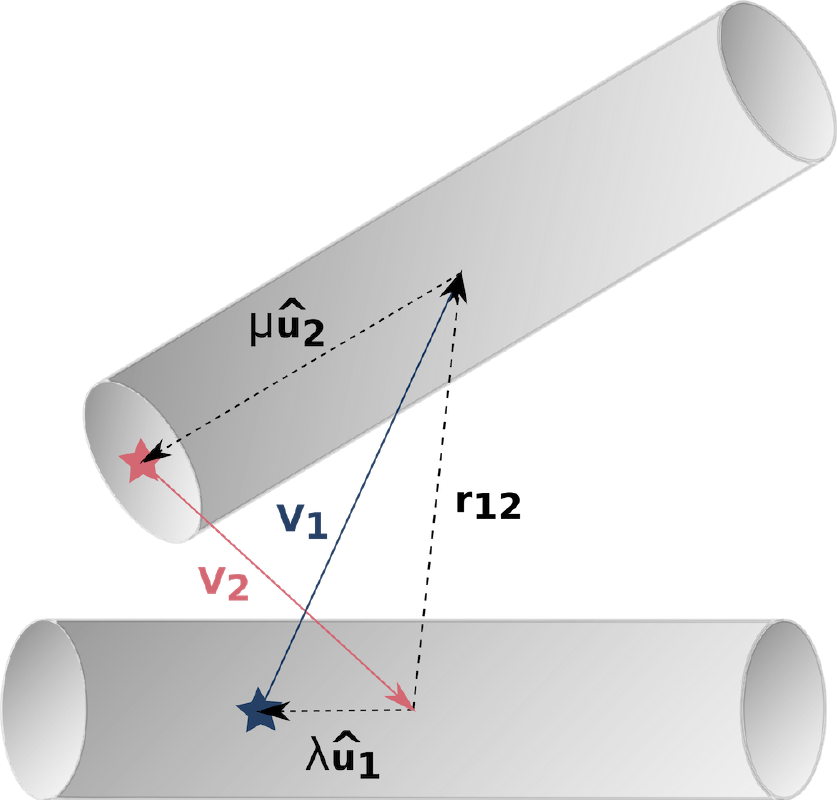}}
  \caption{The star symbols represent the
  points of closest approach on each cylinder.}
  \label{rrtests}
\end{figure}

In Figure~\ref{rrtests}, we see that in the case of a disk-rim configuration,
 for instance, the projection of $\mathbf{V}_1$ on the direction of 
 $\mathbf{\widehat{u_2}}$ is larger than $L/2$.

\subsubsection{Disk-disk overlap}

\begin{figure}[ht!]
	\centering
	\includegraphics[width=0.65\linewidth]{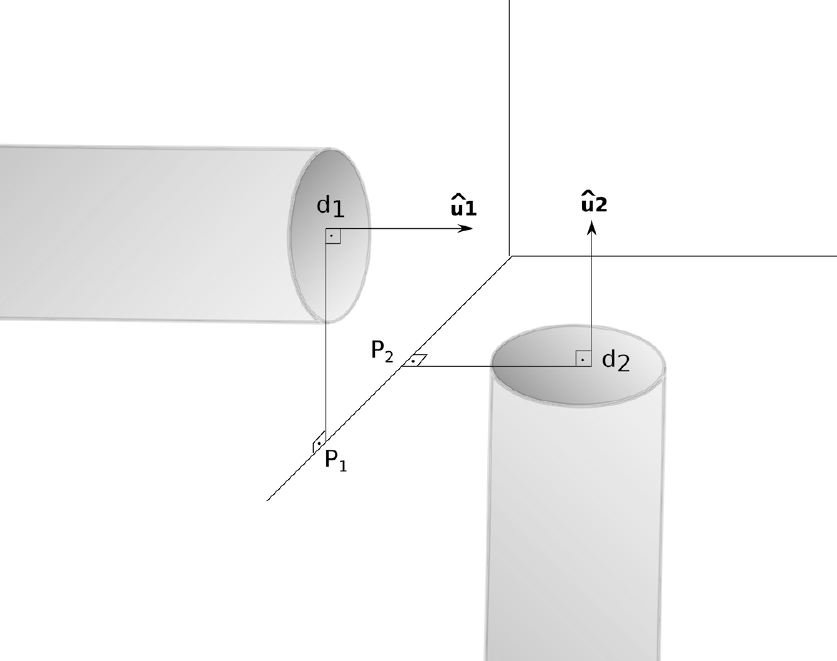}
	\caption{Disks of two cylinders.}
	\label{disks1}
\end{figure}

The orientations of the cylinders are perpendicular to the planes of 
the disks. The planes of the two disks intersect in a line parallel
 to $\mathbf{\widehat{u}_1} \times \mathbf{\widehat{u}}_2$. 
 We define $\mathbf{P}_1$ and $\mathbf{P}_2$ as being the 
points in the intersection line that are closer to the disks centers $\mathbf{d}_1$ 
and $\mathbf{d}_2$, respectively, as shown in Figure~\ref{disks1}. 

To find $\mathbf{P}_1$, we minimise $(\mathbf{P}_1-\mathbf{d}_2)^2$, 
which is equivalent to minimising $|\mathbf{P}_1-\mathbf{d}_1|$.  
The minimisation can be done by applying Lagrange multipliers 
with two constraints: 
\begin{subequations}
\begin{equation}
	\left(\mathbf{P}_1 - \mathbf{d}_1\right)\cdot \mathbf{\widehat{u}_1} = 0
\label{c1}
\end{equation}
\begin{equation}
	\left(\mathbf{P}_1 - \mathbf{d}_2\right)\cdot \mathbf{\widehat{u}_2} = 0
\label{c2}
\end{equation}
\label{constraints}
\end{subequations}

The constraints presented in Equation~\ref{constraints} ensure that $\mathbf{P}_1$ is 
in a line perpendicular to both $\mathbf{\widehat{u}_1}$ and $\mathbf{\widehat{u}_2}$. Applying the Lagrange multipliers:
\begin{equation}
\mathcal{L} = \left(\mathbf{P}_1 - \mathbf{d}_1\right)^2 - \lambda \left(\mathbf{P}_1 - \mathbf{d}_1\right)\cdot \mathbf{\widehat{u}_1} - 
\mu (\mathbf{P}_1 - \mathbf{d}_2) \cdot \mathbf{\widehat{u}_2}
\end{equation}

From $\nabla \mathcal{L} = 0$, one has:
\begin{equation}
\mathbf{P}_1 = \mathbf{d}_1 + \frac{\lambda \mathbf{\widehat{u}_1}}{2} + \frac{\mu \mathbf{\widehat{u}_2}}{2}
\label{p1der}
\end{equation}

Replacing Equation~\ref{c1} into Equation~\ref{p1der}:
\begin{equation}
\lambda = - \mu \left(\mathbf{\widehat{u}_1} \cdot \mathbf{\widehat{u}_2}\right)
\label{lamb1}
\end{equation}

Substituting Equations~\ref{c2} and \ref{lamb1} into \ref{p1der} yields:
\begin{equation}
 \mu = \frac{-2\left(\mathbf{d}_1 - \mathbf{d}_2\right)\cdot \mathbf{\widehat{u}_2}}{1 - \left(\mathbf{\widehat{u}_1} \cdot \mathbf{\widehat{u}_2})\right)^2}
\label{mu1}
\end{equation}

Replacing Equation~\ref{mu1} into \ref{lamb1}:
\begin{equation}
\lambda = \frac{2\left[\left(\mathbf{d}_1 - \mathbf{d}_2\right)\cdot \mathbf{\widehat{u}_2}\right]\cdot(\mathbf{\widehat{u}_1}\cdot\mathbf{\widehat{u}_2})}
{1 - (\mathbf{\widehat{u}_1} \cdot \mathbf{\widehat{u}_2})^2}
\label{lamb2}
\end{equation}

Replacing Equations~\ref{lamb2} and \ref{mu1} into \ref{p1der}:
\begin{equation}
\mathbf{P}_1 = \mathbf{d}_1 + \frac{\left[\left(d_1 - d_2\right)\cdot \mathbf{\widehat{u}_2}\right] \cdot
 \left(\left(\mathbf{\widehat{u}_1} \cdot \mathbf{\widehat{u}_2}\right) \cdot \mathbf{\widehat{u}_1} -  \mathbf{\widehat{u}_2}\right)}
{1 - \left(\mathbf{\widehat{u}_1} \cdot \mathbf{\widehat{u}_2}\right)^2}
\label{p11}
\end{equation}

We define $\mathbf{d_{12}} = \mathbf{d}_2 - \mathbf{d}_1 $ and $\Delta_{1}^2 = (\mathbf{P}_1 - \mathbf{d}_1)^2$, and rewrite
 Equation~\ref{p11} as:
\begin{equation}
\Delta_1^2 = \frac{(\mathbf{d_{12}}\cdot\mathbf{\widehat{u}_2})^2\cdot 
((\mathbf{\widehat{u}_1} \cdot \mathbf{\widehat{u}_2})^2 - 2(\mathbf{\widehat{u}_1}\cdot \mathbf{\widehat{u}_2})^2 + 1)}
{(1 - (\mathbf{\widehat{u}_1} \cdot \mathbf{\widehat{u}_2})^2)^2}
\label{dist1}
\end{equation}
 
Simplifying Equation~\ref{dist1}:
\begin{equation}
\Delta_1^2 = \frac{(\mathbf{d_{12}}\cdot\mathbf{\widehat{u}_2})^2} 
{1 - (\mathbf{\widehat{u}_1} \cdot \mathbf{\widehat{u}_2})^2}
\end{equation}

Similarly for disk 2:
\begin{equation}
\Delta_2^2 = \frac{(\mathbf{d_{12}}\cdot\mathbf{\widehat{u}_1})^2} 
{1 - (\mathbf{\widehat{u}_1} \cdot \mathbf{\widehat{u}_2})^2}
\end{equation}

\begin{figure}[ht!]
	\centering
	\includegraphics[width=0.65\linewidth]{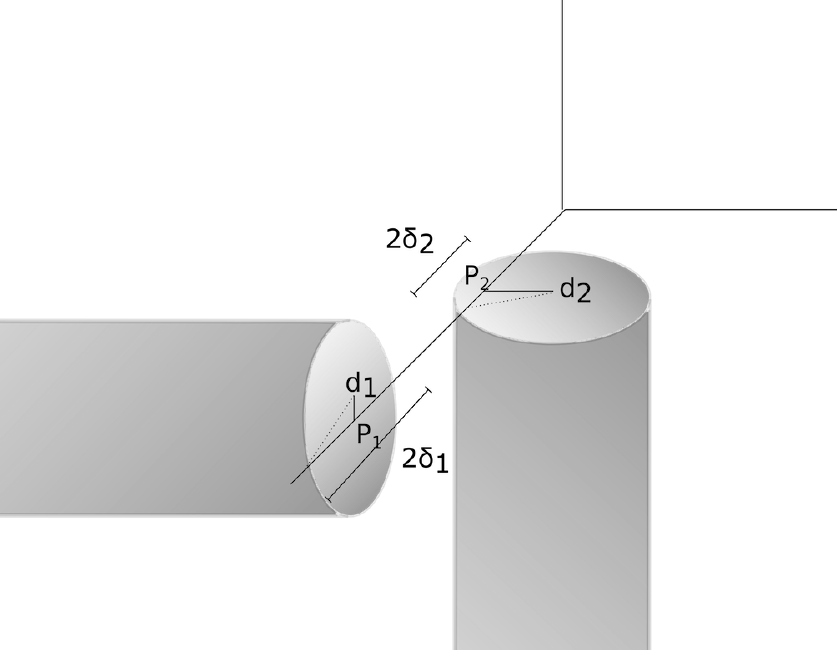}
	\caption{Disks of two cylinders.}
	\label{disks2}
\end{figure}

A necessary, but not sufficient, condition for the overlap to occur is that 
both $\Delta_1$ and $\Delta_2$ have to be less than the cylinder radius $D/2$. If this condition is satisfied, the intersection line crosses both disks through segments of length $2\delta_1$ and $2\delta_2$, as presented in Figure~\ref{disks2}.

The expressions to calculate $\delta_1$ and $\delta_2$ are presented in Equation~\ref{segments}.
 \begin{equation}
	 \begin{aligned}
		 \delta_1 &= \sqrt{\frac{D^2}{4} - \Delta_1^2} \\
		 \delta_2 &= \sqrt{\frac{D^2}{4} - \Delta_2^2}
	 \end{aligned}
	 \label{segments}
 \end{equation}

Finally, an overlap will occur if the condition in the following equation is true:
 \begin{equation}
	 \left|\mathbf{P}_2 - \mathbf{P}_1\right| = 
	\left|\mathbf{d_{12}}\cdot \frac{(\mathbf{\widehat{u}_1} \times \mathbf{\widehat{u}_1})}
	 {|\mathbf{\widehat{u}_1} \times \mathbf{\widehat{u}_2} |}\right|
	 \leq \delta_1 + \delta_2
	 \label{ddovp}
 \end{equation}

\subsubsection{Disk-rim overlap}

Let us take a disk with centre in $\mathbf{d}_j$ and a cylinder with centre in $\mathbf{r}_i$. We define 
$\mathbf{U}_i$ as the point on cylinder $i$ that is the closest to $\mathbf{d}_j$, $\mathbf{P}_d$ a point on the disk $j$ that is the closest to cylinder $i$, $\mathbf{P}_c$ a point on cylinder $i$ that is the closest to disk $j$, $\phi$ an angle between $\widehat{\mathbf{w}}_{j}$ and $\mathbf{d}_{j}-\mathbf{P}_{d}$, $\widehat{\mathbf{v}}_{j}$, $\widehat{\mathbf{u}}_{j}$, an axis system fixed 
on cylinder $j$ and, finally, $\phi$ as an angle between $\widehat{\mathbf{w}}_{j}$ and $\mathbf{d}_{j}- \mathbf{P}_{d}$.    
\begin{figure}[ht!]
	\centering
	\includegraphics[width=0.75\linewidth]{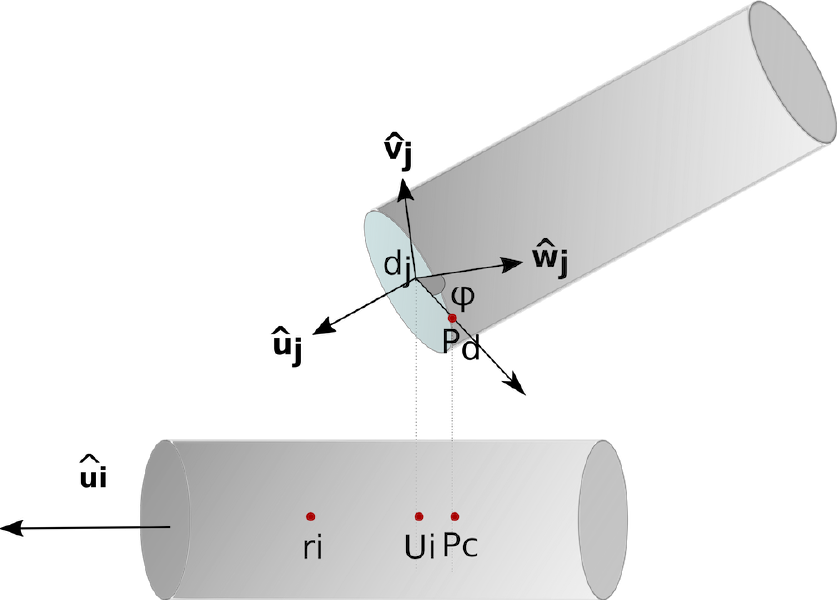}
	\caption{Disk-rim configuration.}
	\label{disk_rim}
\end{figure}

$\mathbf{U}_i$ is obtained from:

\begin{equation}
	\mathbf{U}_i = \mathbf{r}_i + \left[\left(\mathbf{d}_j - \mathbf{r}_i\right)\cdot \widehat{\mathbf{u}}_{i}\right]\widehat{\mathbf{u}}_{i}
\end{equation}

First, we test the following conditions:
\begin{enumerate}
	\item If $|\mathbf{d}_j - \mathbf{U}_i| > d$ : there is no overlap 
	\item If $|\mathbf{d}_j - \mathbf{U}_i| < d/2$ and $|\mathbf{d}_j - \mathbf{r}_i| > L/2$ : the overlap would be 
a disk-disk kind and not a disk-rim, and therefore we do not need to handle this condition test at this stage.
	\item If $|\mathbf{d}_j - \mathbf{U}_i| \leq d/2$ and $|(\mathbf{d}_j - \mathbf{r}_i)| < L/2$ : the two cylinders are overlapping, since the centre of the disk $j$ is within cylinder $i$.
\end{enumerate}

Test number 3 is a sufficient, but not necessary, condition for the overlap 
to occur, since another point can be touching cylinder $j$ even if 
$\mathbf{d}_j$ is not within cylinder $i$. 

Hence, if condition 3 is not satisfied, we have to find $\mathbf{P}_d$, the closest point in disk $j$ to cylinder $i$.

Arbitrary points on the border of disk $j$ ($\mathbf{d}$), and on the line of cylinder $i$ ($\mathbf{c}$) are defined as:
\begin{subequations}
	\begin{equation}
		\mathbf{d} = \mathbf{d}_j + R\cos{(\phi)}\widehat{\mathbf{w}}_{j} +
			R\sin{(\phi)}\widehat{\mathbf{v}}_{j}
	\end{equation}
	\begin{equation}
		\mathbf{c} = \mathbf{r}_i + \lambda\mathbf{\widehat{u}_i}
	\end{equation}
\end{subequations}
where $R \equiv D/2$ is the radius of the cylinders. 

The square of the distance between $\mathbf{d}$ and $\mathbf{c}$ is thus:

\begin{eqnarray}
	\label{d-c}
\left(\mathbf{d}-\mathbf{c}\right)^2 &=& \left(\mathbf{d}_{j} -\mathbf{r}_{2}\right)^2  + R^2 + \lambda^2 \\ \nonumber 
&+&	2R\cos{\phi}\left(\left(\mathbf{d}_j-\mathbf{r}_{i}\right) \cdot \widehat{\mathbf{w}}_{j}\right)	\\ \nonumber
&+& 2R\sin{\phi}\left(\left(\mathbf{d}_{j}\mathbf{r}_{i}\right) \cdot \widehat{\mathbf{v}}_{j}\right) - 2\lambda\left(\left(\mathbf{d}_{j}-\mathbf{r}_{i}\right)\cdot \widehat{\mathbf{u}}_{i}\right) \\ \nonumber
	&-& 2\lambda R \cos{\phi}\left(\widehat{\mathbf{w}}_{j}\cdot \widehat{\mathbf{u}}_{i}\right)
	- 2\lambda R \sin{\phi}\left(\widehat{\mathbf{v}}_{j}\cdot \widehat{\mathbf{u}}_{i}\right) 
\end{eqnarray}

$\mathbf{P}_c$ and $\mathbf{P}_d$ are the points that minimise Equation~\ref{d-c}, therefore:

\begin{eqnarray}
		\lambda &-& r\cos{\phi}(\mathbf{\widehat{w}_{j}} \cdot \mathbf{\widehat{u}_i})
		- r\sin{\phi}(\mathbf{\widehat{v}_j}\cdot \mathbf{\widehat{u}_i}) + \\ \nonumber
		&-& \left(\left(\mathbf{d}_{j}-\mathbf{r}_{i}\right)\cdot \widehat{\mathbf{u}}_{i}\right) = 0
		\label{dgdl}
	\end{eqnarray}
	\begin{eqnarray}
		&&\sin{\phi}\left[\lambda\left(\widehat{\mathbf{w}}_{j}\cdot \widehat{\mathbf{u}}_{i}\right) - 
		\left(\left(\mathbf{d}_{j}-\mathbf{r}_{i}\right)\cdot \widehat{\mathbf{w}}_{j}\right)\right] + \\ \nonumber
		&-& \cos{\phi}\left[\lambda\left(\widehat{\mathbf{v}}_{j}\cdot \widehat{\mathbf{u}}_{i}\right) 
		- \left(\left(\mathbf{d}_{j}-\mathbf{r}_{i}\right)\cdot \widehat{\mathbf{v}}_{j}\right)\right] = 0
		\label{dgdp}
	\end{eqnarray}

Rewriting Equation~\ref{dgdp} gives:
\begin{eqnarray}
	\label{sincos}
	\frac{\sin{\phi}}{\cos{\phi}} = 
		\frac{\lambda\left(\widehat{\mathbf{v}}_{j}\cdot \widehat{\mathbf{u}}_{i}\right) 
		- \left(\left(\mathbf{d}_{j}-\mathbf{r}_{i}\right)\cdot \widehat{\mathbf{v}}_{j}\right)}
		{\lambda\left(\widehat{\mathbf{w}}_{j}\cdot \widehat{\mathbf{u}}_{i}\right) - 
		\left(\left(\mathbf{d}_{j}-\mathbf{r}_{i}\right)\cdot \widehat{\mathbf{w}}_{j}\right)}
\end{eqnarray}

If the numerator and denominator of Equation~\ref{sincos} are 
taken as the catheti of a triangle, the hypotenuse can then be found 
to give the expressions for $\cos{\phi}$ and $\sin{\phi}$. Once we have 
these expressions, they are applied into Equation~\ref{dgdl}, resulting 
in an equation for $\lambda$. Since we were not able to find an analytical solution to the previous equation, a numerical method such as the Newton-Raphson or bisebsection method is used to find $\lambda$. In our code, we combine both methods, running a few steps with one and a few with the other until convergence is found to machine precision.

Once $\mathbf{P}_d$ is obtained, we define $\mathbf{T} = \mathbf{P}_d - \mathbf{r}_i$, and calculate the components of $\mathbf{T}$ that are parallel $\mathbf{T}_\parallel$ and perpendicular $\mathbf{T}_\bot$ to $\widehat{\mathbf{u}}_{i}$.
\begin{subequations}
\begin{equation}
	\mathbf{T}_\parallel = \left(\mathbf{T} \cdot \widehat{\mathbf{u}}_{1}\right)\widehat{\mathbf{u}}_{1}
\end{equation}
\begin{equation}
	\mathbf{T}_\bot= \mathbf{T} - \mathbf{T}_\parallel
\end{equation}
\end{subequations}

Finally, the overlap only occurs if $|\mathbf{T}_\parallel| \le L/2$ and $|\mathbf{T}_\bot| \le D/2 $.
\section{Supplementary material}
\label{sec:supplementary}
\subsection{Figures}
\begin{figure}[ht!]
    \centering
    \subfigure[]{\includegraphics[width = 0.45\textwidth]{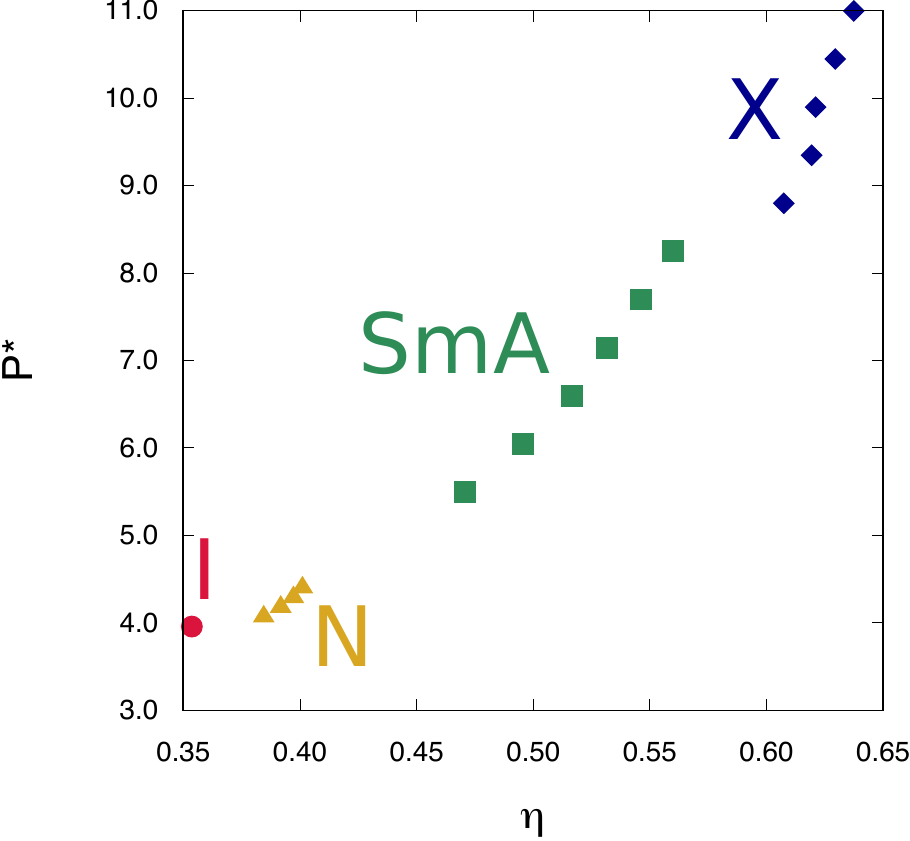}}  
    \subfigure[]{\includegraphics[width = 0.45\textwidth]{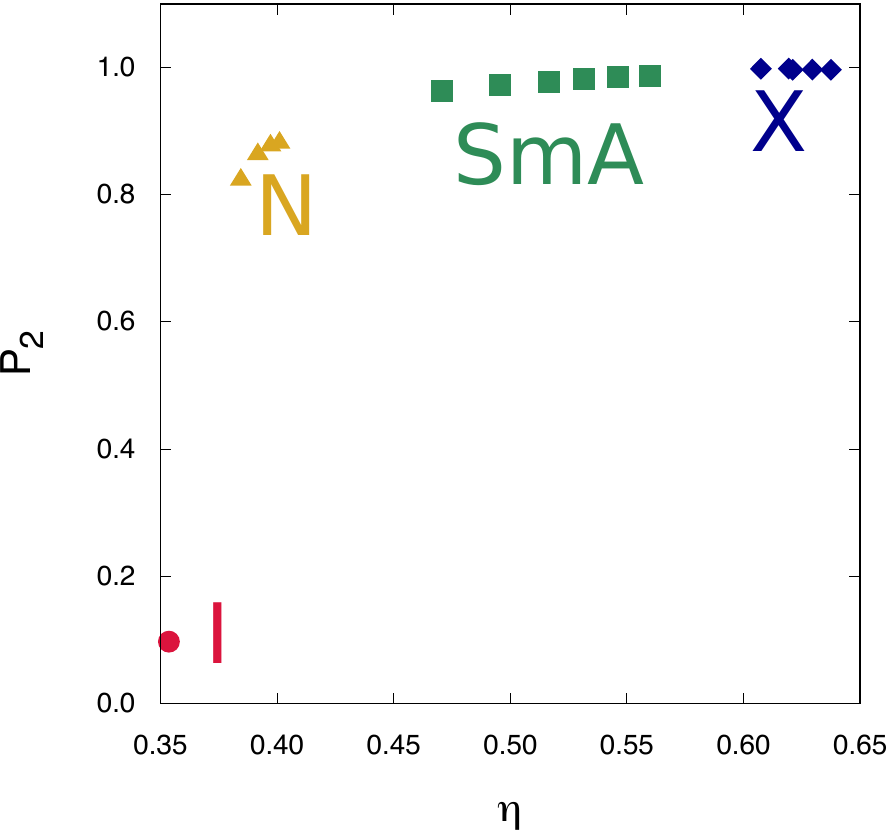}}   
     \caption{(a) Reduced pressure $P^{*}$ versus volume fraction $\eta$ for $L/D = 7$. (b) Nematic order parameter $P_2$ versus volume fraction $\eta$ for $L/D=7$.}
     \label{fig:EOS_cylinders_l7}
    \end{figure}          
\begin{figure}[ht!]
    \centering
    \subfigure[]{\includegraphics[width = 0.55\textwidth]{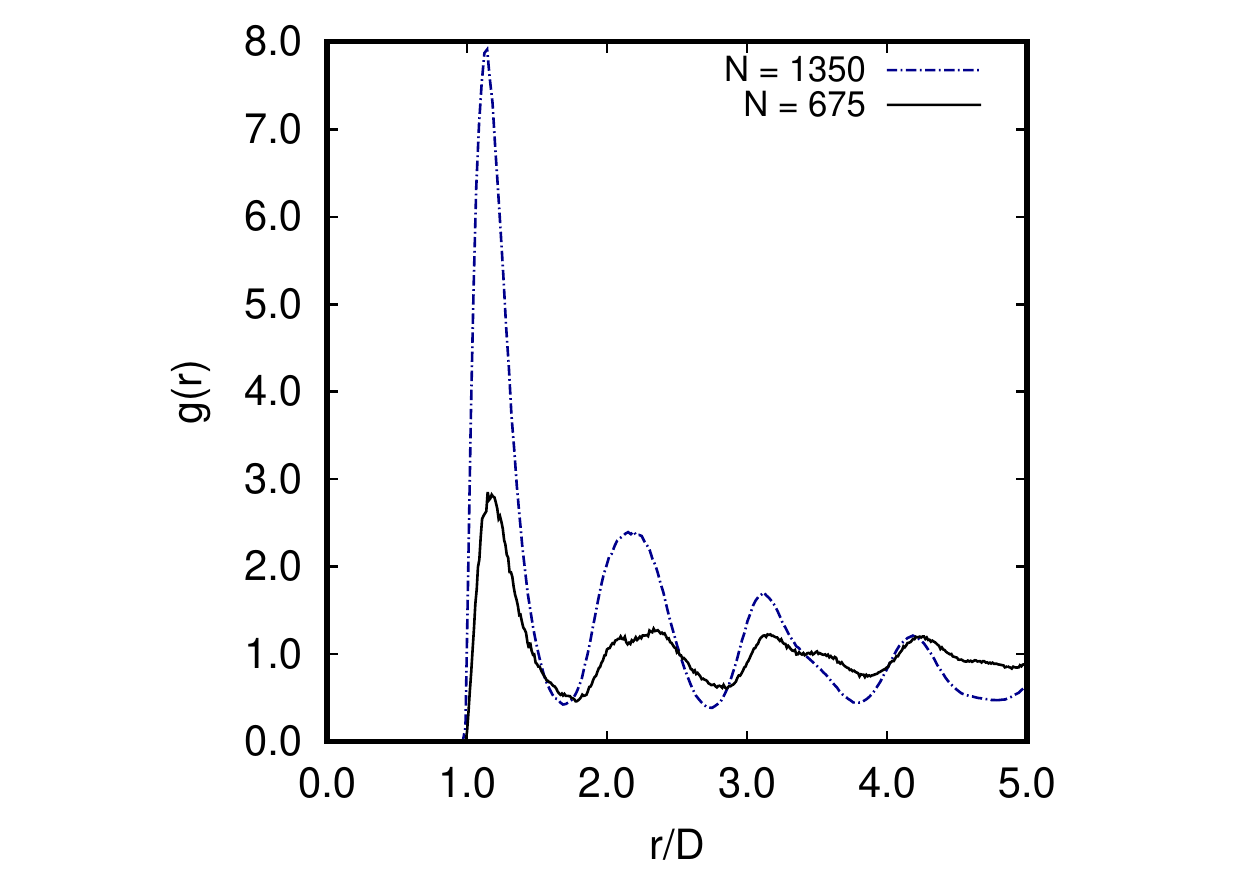}} 
    \subfigure[]{\includegraphics[width = 0.55\textwidth]{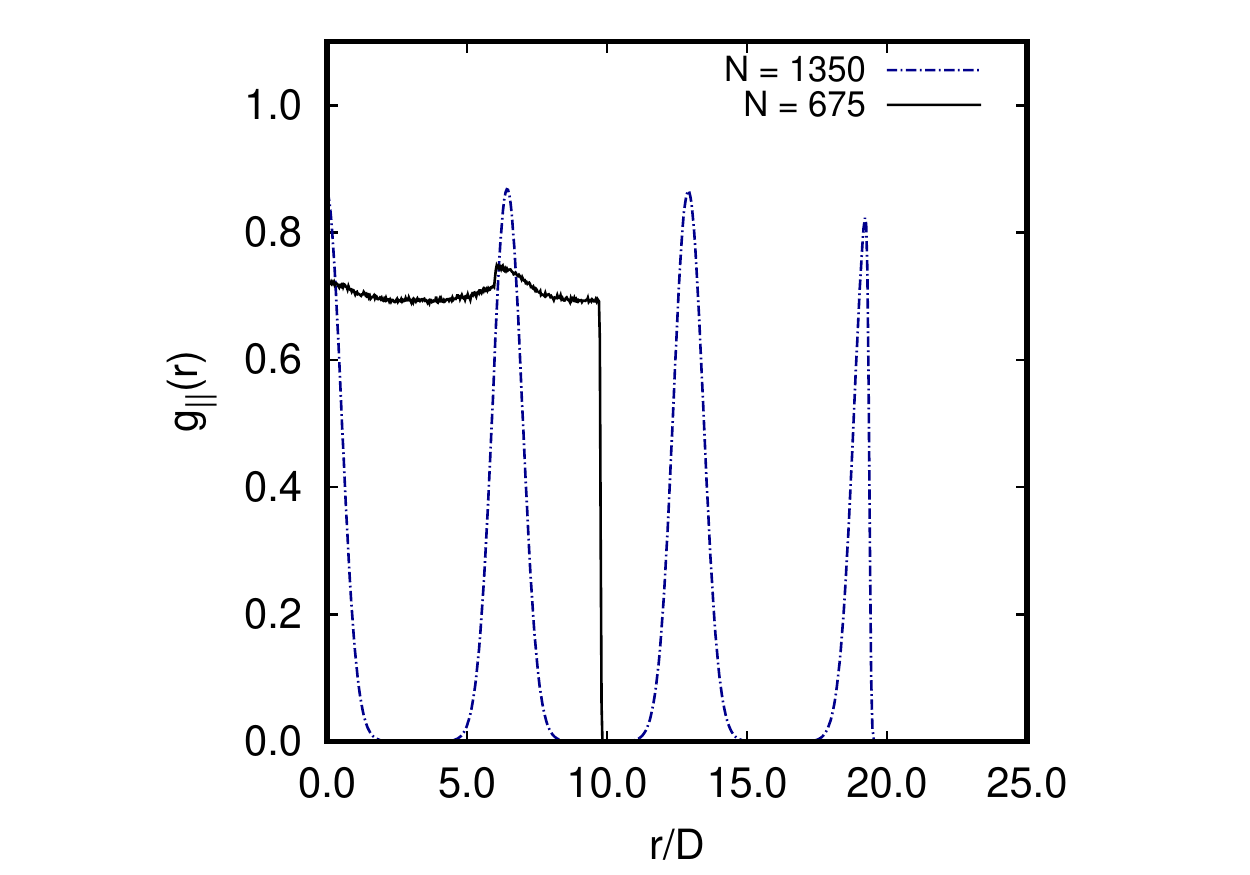}}
     \subfigure[]{\includegraphics[width =0.55\textwidth]{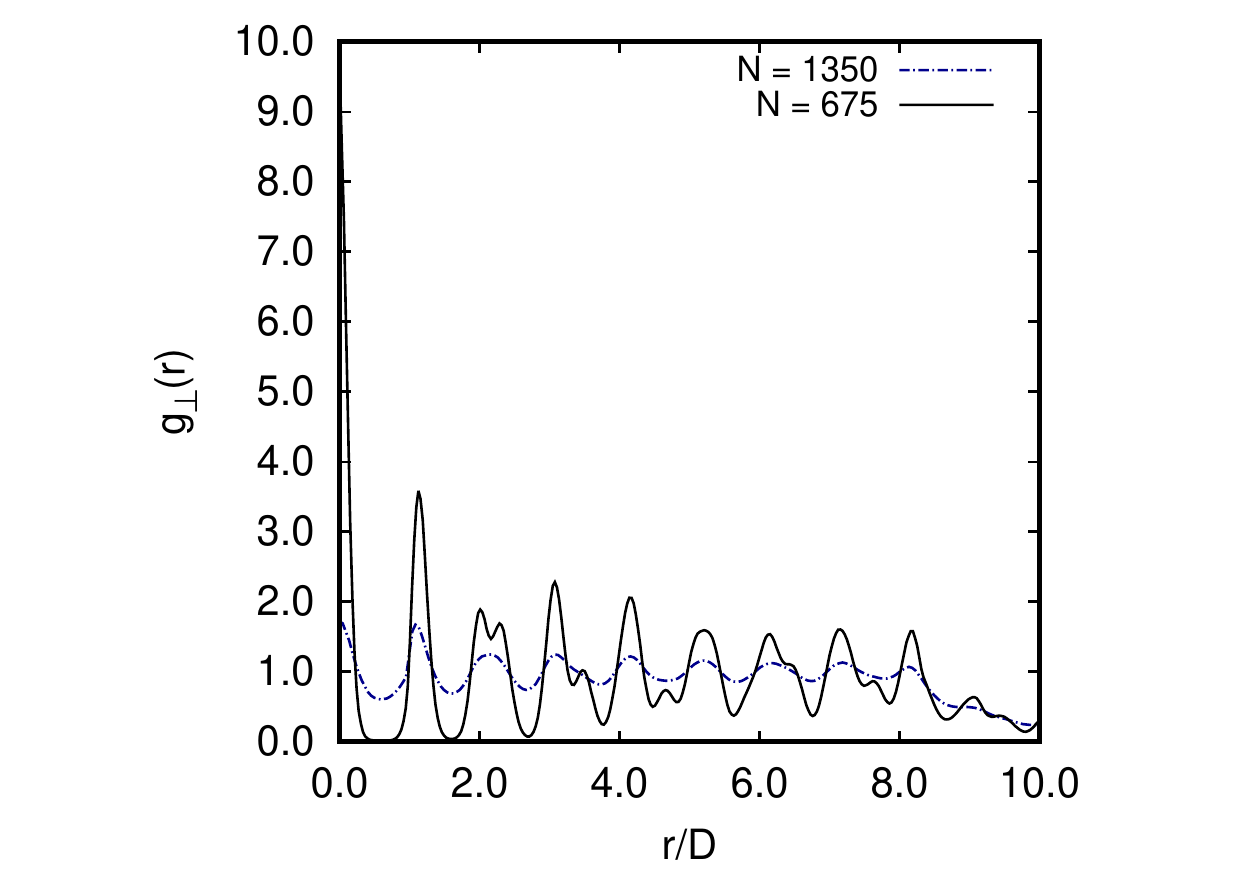}}
     \caption{Distribution functions of hard cylinders with aspect ratio $L/D = 6$  and $\eta = 0.6$x§
      (a) $g(r)$; (b) $g_{\parallel}(r_{\parallel})$; (c) $g_{\perp}(r_{\perp})$. Dotted line $N=1350$ and $P^*=9.42$ , Solid line $N=675$ and $P^*=8.01$.}
     \label{fig:df}
    \end{figure}          
\begin{figure}[ht!]
    \centering
    \subfigure[]{\includegraphics[width = 0.45\textwidth]{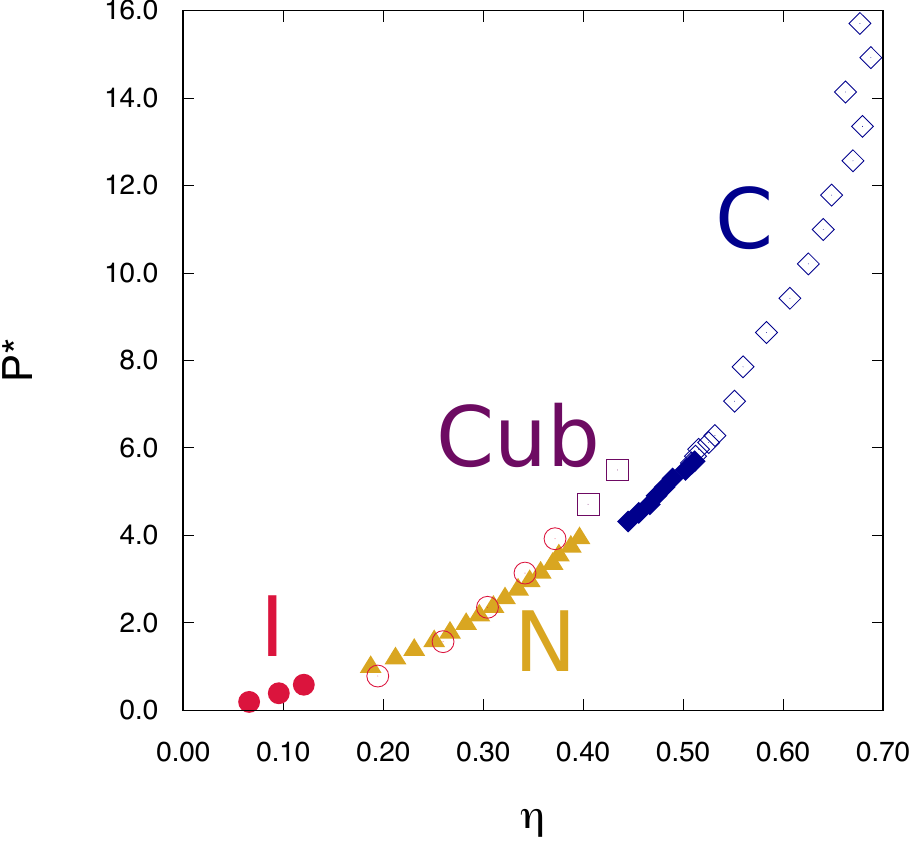}}  
    \subfigure[]{\includegraphics[width = 0.45\textwidth]{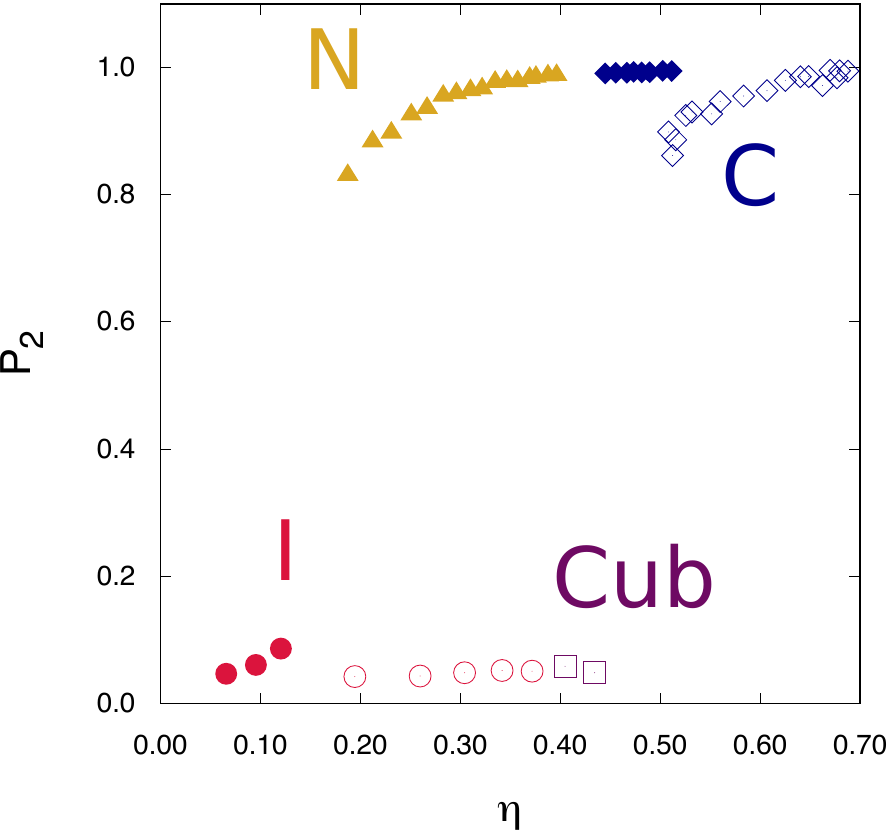}}
     \caption{(a) Reduced pressure $P^{*}$ versus volume fraction $\eta$. Open symbols: $L/D = 0.2$, closed symbols: $L/D = 0.05$; (b) Nematic order parameter $P_2$ versus volume fraction $\eta$ for both $L/D=0.2$ and $L/D=0.05$, with the same symbols as before.}
     \label{fig:EOS_disks}
    \end{figure}          


\clearpage
\subsection{Tables}
 \begin{table}[h!]
	\caption{$L/D= 0.05 $}
         \centering
        \footnotesize
        \label{ld0.05}
	\begin{tabular}{c c c c c c}
	\hline
	$P^*$ &  	 $\eta$ &  	  $S$ &   	 $\psi_6$ &	 $\tau$ & 	 Phase \\ \hline
0.20	 &	  0.066 $\pm$ 0.001	 &	 0.047 $\pm$ 0.008	 &	 0.068 $\pm$ 0.007	 &	 0.178 $\pm$ 0.052	 &	I	 \\ 
0.39	 &	  0.096 $\pm$ 0.001	 &	 0.061 $\pm$ 0.002	 &	 0.133 $\pm$ 0.003	 &	 0.181 $\pm$ 0.049	 &	I	 \\ 
0.59	 &	  0.121 $\pm$ 0.001	 &	 0.086 $\pm$ 0.027	 &	 0.190 $\pm$ 0.006	 &	 0.185 $\pm$ 0.060	 &	I	 \\ 
0.98	 &	  0.187 $\pm$ 0.001	 &	 0.830 $\pm$ 0.019	 &	 0.308 $\pm$ 0.009	 &	 0.253 $\pm$ 0.022	 &	N	 \\ 
1.18	 &	  0.212 $\pm$ 0.003	 &	 0.883 $\pm$ 0.014	 &	 0.332 $\pm$ 0.009	 &	 0.193 $\pm$ 0.013	 &	N	 \\ 
1.37	 &	  0.231 $\pm$ 0.004	 &	 0.896 $\pm$ 0.031	 &	 0.360 $\pm$ 0.011	 &	 0.203 $\pm$ 0.012	 &	N	 \\ 
1.57	 &	  0.251 $\pm$ 0.003	 &	 0.925 $\pm$ 0.008	 &	 0.386 $\pm$ 0.008	 &	 0.178 $\pm$ 0.018	 &	N	 \\ 
1.77	 &	  0.267 $\pm$ 0.001	 &	 0.935 $\pm$ 0.009	 &	 0.405 $\pm$ 0.007	 &	 0.204 $\pm$ 0.012	 &	N	 \\ 
1.96	 &	  0.283 $\pm$ 0.005	 &	 0.955 $\pm$ 0.008	 &	 0.427 $\pm$ 0.006	 &	 0.113 $\pm$ 0.022	 &	N	 \\ 
2.16	 &	  0.296 $\pm$ 0.004	 &	 0.959 $\pm$ 0.005	 &	 0.437 $\pm$ 0.011	 &	 0.205 $\pm$ 0.011	 &	N	 \\ 
2.36	 &	  0.310 $\pm$ 0.003	 &	 0.963 $\pm$ 0.009	 &	 0.447 $\pm$ 0.013	 &	 0.194 $\pm$ 0.015	 &	N	 \\ 
2.55	 &	  0.322 $\pm$ 0.001	 &	 0.966 $\pm$ 0.006	 &	 0.455 $\pm$ 0.010	 &	 0.216 $\pm$ 0.003	 &	N	 \\ 
2.75	 &	  0.335 $\pm$ 0.003	 &	 0.976 $\pm$ 0.005	 &	 0.477 $\pm$ 0.009	 &	 0.176 $\pm$ 0.013	 &	N	 \\ 
2.95	 &	  0.346 $\pm$ 0.004	 &	 0.978 $\pm$ 0.006	 &	 0.483 $\pm$ 0.015	 &	 0.205 $\pm$ 0.008	 &	N	 \\ 
3.14	 &	  0.357 $\pm$ 0.005	 &	 0.977 $\pm$ 0.007	 &	 0.488 $\pm$ 0.010	 &	 0.198 $\pm$ 0.007	 &	N	 \\ 
3.34	 &	  0.369 $\pm$ 0.005	 &	 0.983 $\pm$ 0.004	 &	 0.500 $\pm$ 0.008	 &	 0.142 $\pm$ 0.015	 &	N	 \\ 
3.53	 &	  0.376 $\pm$ 0.003	 &	 0.984 $\pm$ 0.002	 &	 0.501 $\pm$ 0.005	 &	 0.172 $\pm$ 0.016	 &	N	 \\ 
3.73	 &	  0.388 $\pm$ 0.005	 &	 0.986 $\pm$ 0.001	 &	 0.515 $\pm$ 0.009	 &	 0.198 $\pm$ 0.011	 &	N	 \\ 
3.93	 &	  0.396 $\pm$ 0.007	 &	 0.987 $\pm$ 0.003	 &	 0.513 $\pm$ 0.010	 &	 0.158 $\pm$ 0.005	 &	N	 \\ 
4.32	 &	  0.445 $\pm$ 0.007	 &	 0.990 $\pm$ 0.001	 &	 0.586 $\pm$ 0.014	 &	 0.185 $\pm$ 0.007	 &	C	 \\ 
4.52	 &	  0.455 $\pm$ 0.008	 &	 0.991 $\pm$ 0.000	 &	 0.597 $\pm$ 0.017	 &	 0.210 $\pm$ 0.005	 &	C	 \\ 
4.71	 &	  0.467 $\pm$ 0.006	 &	 0.992 $\pm$ 0.000	 &	 0.613 $\pm$ 0.004	 &	 0.189 $\pm$ 0.004	 &	C	 \\ 
4.91	 &	  0.473 $\pm$ 0.006	 &	 0.993 $\pm$ 0.001	 &	 0.607 $\pm$ 0.016	 &	 0.195 $\pm$ 0.014	 &	C	 \\ 
5.11	 &	  0.481 $\pm$ 0.007	 &	 0.992 $\pm$ 0.002	 &	 0.626 $\pm$ 0.017	 &	 0.180 $\pm$ 0.012	 &	C	 \\ 
5.30	 &	  0.489 $\pm$ 0.006	 &	 0.992 $\pm$ 0.002	 &	 0.633 $\pm$ 0.020	 &	 0.200 $\pm$ 0.015	 &	C	 \\ 
5.50	 &	  0.502 $\pm$ 0.006	 &	 0.994 $\pm$ 0.001	 &	 0.643 $\pm$ 0.016	 &	 0.204 $\pm$ 0.014	 &	C	 \\ 
5.69	 &	  0.511 $\pm$ 0.007	 &	 0.994 $\pm$ 0.000	 &	 0.632 $\pm$ 0.012	 &	 0.208 $\pm$ 0.005	 &	C	 \\ 
7.85	 &	  0.575 $\pm$ 0.001	 &	 0.999 $\pm$ 0.000	 &	 0.397 $\pm$ 0.006	 &	 0.217 $\pm$ 0.003	 &	C	 \\ 
	\hline
	\end{tabular}
\end{table}
  \begin{table}[h!]
	\caption{$L/D= 0.20 $}
	\centering
        \footnotesize
\label{ld0.2}
	 \begin{tabular}{c c c c c c}
	\hline
	$P^*$ &  	 $\eta$ &  	  $S$ &   	 $\psi_6$ &	 $\tau$ & 	 Phase \\ \hline
0.79	 &	  0.194 $\pm$ 0.002	 &	 0.042 $\pm$ 0.004	 &	 0.285 $\pm$ 0.008	 &	 0.120 $\pm$ 0.067	 &	I	 \\ 
1.57	 &	  0.260 $\pm$ 0.001	 &	 0.043 $\pm$ 0.006	 &	 0.391 $\pm$ 0.006	 &	 0.130 $\pm$ 0.050	 &	I	 \\ 
2.36	 &	  0.304 $\pm$ 0.002	 &	 0.048 $\pm$ 0.011	 &	 0.434 $\pm$ 0.010	 &	 0.128 $\pm$ 0.049	 &	I	 \\ 
3.14	 &	  0.342 $\pm$ 0.003	 &	 0.052 $\pm$ 0.011	 &	 0.450 $\pm$ 0.004	 &	 0.129 $\pm$ 0.043	 &	I	 \\ 
3.93	 &	  0.372 $\pm$ 0.004	 &	 0.051 $\pm$ 0.012	 &	 0.460 $\pm$ 0.006	 &	 0.114 $\pm$ 0.076	 &	I	 \\ 
4.71	 &	  0.405 $\pm$ 0.004	 &	 0.058 $\pm$ 0.028	 &	 0.458 $\pm$ 0.007	 &	 0.135 $\pm$ 0.029	 &	Cub	 \\ 
5.50	 &	  0.434 $\pm$ 0.004	 &	 0.048 $\pm$ 0.016	 &	 0.455 $\pm$ 0.008	 &	 0.122 $\pm$ 0.062	 &	Cub	 \\ 
6.28	 &	  0.532 $\pm$ 0.003	 &	 0.930 $\pm$ 0.006	 &	 0.767 $\pm$ 0.006	 &	 0.216 $\pm$ 0.004	 &	C	 \\ 
7.07	 &	  0.551 $\pm$ 0.004	 &	 0.927 $\pm$ 0.007	 &	 0.788 $\pm$ 0.010	 &	 0.207 $\pm$ 0.011	 &	C	 \\ 
7.85	 &	  0.560 $\pm$ 0.001	 &	 0.946 $\pm$ 0.004	 &	 0.808 $\pm$ 0.006	 &	 0.218 $\pm$ 0.002	 &	C	 \\ 
8.64	 &	  0.583 $\pm$ 0.004	 &	 0.955 $\pm$ 0.006	 &	 0.822 $\pm$ 0.008	 &	 0.183 $\pm$ 0.008	 &	C	 \\ 
9.42	 &	  0.607 $\pm$ 0.006	 &	 0.963 $\pm$ 0.006	 &	 0.837 $\pm$ 0.011	 &	 0.110 $\pm$ 0.008	 &	C	 \\ 
10.21	 &	  0.625 $\pm$ 0.004	 &	 0.979 $\pm$ 0.003	 &	 0.840 $\pm$ 0.010	 &	 0.099 $\pm$ 0.015	 &	C	 \\ 
11.00	 &	  0.640 $\pm$ 0.003	 &	 0.985 $\pm$ 0.002	 &	 0.840 $\pm$ 0.008	 &	 0.111 $\pm$ 0.006	 &	C	 \\ 
11.78	 &	  0.648 $\pm$ 0.002	 &	 0.986 $\pm$ 0.001	 &	 0.856 $\pm$ 0.007	 &	 0.132 $\pm$ 0.003	 &	C	 \\ 
12.57	 &	  0.670 $\pm$ 0.003	 &	 0.995 $\pm$ 0.001	 &	 0.726 $\pm$ 0.010	 &	 0.210 $\pm$ 0.009	 &	C	 \\ 
13.35	 &	  0.679 $\pm$ 0.003	 &	 0.994 $\pm$ 0.001	 &	 0.795 $\pm$ 0.008	 &	 0.156 $\pm$ 0.005	 &	C	 \\ 
14.14	 &	  0.662 $\pm$ 0.002	 &	 0.971 $\pm$ 0.002	 &	 0.658 $\pm$ 0.012	 &	 0.202 $\pm$ 0.002	 &	C	 \\ 
14.92	 &	  0.688 $\pm$ 0.002	 &	 0.994 $\pm$ 0.000	 &	 0.858 $\pm$ 0.007	 &	 0.135 $\pm$ 0.006	 &	C	 \\ 
15.71	 &	  0.677 $\pm$ 0.002	 &	 0.983 $\pm$ 0.000	 &	 0.745 $\pm$ 0.009	 &	 0.187 $\pm$ 0.006	 &	C	 \\ 
	\hline
	\end{tabular}
\end{table}
   \begin{table}[h!]
	\caption{$L/D= 5.00 $}
	\centering
        \footnotesize
\label{ld5}
	 \begin{tabular}{c c c c c c}
	\hline
	$P^*$ &  	 $\eta$ &  	  $S$ &   	 $\psi_6$ &	 $\tau$ & 	 Phase \\ \hline
0.79	 &	  0.193 $\pm$ 0.003	 &	 0.029 $\pm$ 0.005	 &	 0.002 $\pm$ 0.001	 &	 0.395 $\pm$ 0.100	 &	I	 \\ 
1.18	 &	  0.229 $\pm$ 0.004	 &	 0.030 $\pm$ 0.004	 &	 0.004 $\pm$ 0.001	 &	 0.367 $\pm$ 0.128	 &	I	 \\ 
1.57	 &	  0.256 $\pm$ 0.004	 &	 0.029 $\pm$ 0.002	 &	 0.007 $\pm$ 0.002	 &	 0.361 $\pm$ 0.124	 &	I	 \\ 
1.96	 &	  0.280 $\pm$ 0.002	 &	 0.034 $\pm$ 0.009	 &	 0.010 $\pm$ 0.002	 &	 0.426 $\pm$ 0.153	 &	I	 \\ 
2.36	 &	  0.299 $\pm$ 0.003	 &	 0.034 $\pm$ 0.007	 &	 0.015 $\pm$ 0.001	 &	 0.350 $\pm$ 0.153	 &	I	 \\ 
2.75	 &	  0.317 $\pm$ 0.004	 &	 0.033 $\pm$ 0.004	 &	 0.021 $\pm$ 0.002	 &	 0.423 $\pm$ 0.149	 &	I	 \\ 
3.14	 &	  0.332 $\pm$ 0.003	 &	 0.032 $\pm$ 0.010	 &	 0.028 $\pm$ 0.004	 &	 0.350 $\pm$ 0.133	 &	I	 \\ 
3.53	 &	  0.348 $\pm$ 0.002	 &	 0.041 $\pm$ 0.006	 &	 0.036 $\pm$ 0.002	 &	 0.335 $\pm$ 0.192	 &	I	 \\ 
3.93	 &	  0.359 $\pm$ 0.002	 &	 0.040 $\pm$ 0.029	 &	 0.045 $\pm$ 0.002	 &	 0.297 $\pm$ 0.222	 &	I	 \\ 
4.32	 &	  0.372 $\pm$ 0.001	 &	 0.045 $\pm$ 0.010	 &	 0.056 $\pm$ 0.003	 &	 0.406 $\pm$ 0.229	 &	I	 \\ 
4.71	 &	  0.385 $\pm$ 0.002	 &	 0.042 $\pm$ 0.010	 &	 0.071 $\pm$ 0.009	 &	 0.270 $\pm$ 0.377	 &	I	 \\ 
5.11	 &	  0.397 $\pm$ 0.002	 &	 0.059 $\pm$ 0.011	 &	 0.089 $\pm$ 0.003	 &	 0.208 $\pm$ 0.098	 &	I	 \\ 
5.50	 &	  0.408 $\pm$ 0.003	 &	 0.044 $\pm$ 0.015	 &	 0.118 $\pm$ 0.011	 &	 0.408 $\pm$ 0.218	 &	I	 \\ 
5.89	 &	  0.480 $\pm$ 0.004	 &	 0.927 $\pm$ 0.002	 &	 0.444 $\pm$ 0.017	 &	 0.625 $\pm$ 0.077	 &	SmA 	 \\ 
6.28	 &	  0.500 $\pm$ 0.002	 &	 0.951 $\pm$ 0.003	 &	 0.504 $\pm$ 0.008	 &	 0.756 $\pm$ 0.088	 &	SmA 	 \\ 
6.68	 &	  0.515 $\pm$ 0.002	 &	 0.963 $\pm$ 0.001	 &	 0.536 $\pm$ 0.008	 &	 0.825 $\pm$ 0.039	 &	SmA 	 \\ 
7.07	 &	  0.526 $\pm$ 0.001	 &	 0.966 $\pm$ 0.002	 &	 0.554 $\pm$ 0.006	 &	 0.839 $\pm$ 0.010	 &	SmA 	 \\ 
7.46	 &	  0.537 $\pm$ 0.003	 &	 0.971 $\pm$ 0.001	 &	 0.556 $\pm$ 0.004	 &	 0.875 $\pm$ 0.012	 &	SmA 	 \\ 
7.85	 &	  0.549 $\pm$ 0.002	 &	 0.974 $\pm$ 0.001	 &	 0.554 $\pm$ 0.003	 &	 0.861 $\pm$ 0.042	 &	SmA 	 \\ 
8.25	 &	  0.561 $\pm$ 0.004	 &	 0.978 $\pm$ 0.001	 &	 0.546 $\pm$ 0.003	 &	 0.880 $\pm$ 0.025	 &	SmA 	 \\ 
8.64	 &	  0.567 $\pm$ 0.002	 &	 0.979 $\pm$ 0.001	 &	 0.539 $\pm$ 0.004	 &	 0.890 $\pm$ 0.032	 &	SmA 	 \\ 
9.03	 &	  0.603 $\pm$ 0.004	 &	 0.992 $\pm$ 0.001	 &	 0.658 $\pm$ 0.010	 &	 0.855 $\pm$ 0.022	 &	X	 \\ 
9.42	 &	  0.608 $\pm$ 0.005	 &	 0.991 $\pm$ 0.001	 &	 0.638 $\pm$ 0.005	 &	 0.895 $\pm$ 0.016	 &	X	 \\ 
9.82	 &	  0.617 $\pm$ 0.003	 &	 0.992 $\pm$ 0.001	 &	 0.649 $\pm$ 0.006	 &	 0.888 $\pm$ 0.011	 &	X	 \\ 
10.21	 &	  0.618 $\pm$ 0.003	 &	 0.991 $\pm$ 0.001	 &	 0.628 $\pm$ 0.006	 &	 0.921 $\pm$ 0.010	 &	X	 \\ 
10.60	 &	  0.633 $\pm$ 0.006	 &	 0.993 $\pm$ 0.001	 &	 0.673 $\pm$ 0.011	 &	 0.897 $\pm$ 0.034	 &	X	 \\ 
11.39	 &	  0.642 $\pm$ 0.004	 &	 0.995 $\pm$ 0.000	 &	 0.687 $\pm$ 0.004	 &	 0.898 $\pm$ 0.010	 &	X	 \\ 
11.78	 &	  0.650 $\pm$ 0.001	 &	 0.995 $\pm$ 0.001	 &	 0.709 $\pm$ 0.009	 &	 0.900 $\pm$ 0.005	 &	X	 \\ 
	\hline
	\end{tabular}
\end{table}

    \begin{table}[h!]
	\caption{$L/D= 7.00 $}
	\centering
        \footnotesize
\label{ld7}
	 \begin{tabular}{c c c c c c}
	\hline
	$P^*$ &  	 $\eta$ &  	  $S$ &   	 $\psi_6$ &	 $\tau$ & 	 Phase \\ \hline
3.96	 &	  0.354 $\pm$ 0.002	 &	 0.097 $\pm$ 0.031	 &	 0.043 $\pm$ 0.005	 &	 0.537 $\pm$ 0.347	 &	I 	 \\ 
4.07	 &	  0.384 $\pm$ 0.003	 &	 0.823 $\pm$ 0.015	 &	 0.060 $\pm$ 0.004	 &	 0.203 $\pm$ 0.005	 &	N	 \\ 
4.18	 &	  0.392 $\pm$ 0.003	 &	 0.863 $\pm$ 0.016	 &	 0.065 $\pm$ 0.005	 &	 0.210 $\pm$ 0.007	 &	N	 \\ 
4.29	 &	  0.397 $\pm$ 0.001	 &	 0.876 $\pm$ 0.009	 &	 0.071 $\pm$ 0.012	 &	 0.198 $\pm$ 0.008	 &	N	 \\ 
4.40	 &	  0.401 $\pm$ 0.002	 &	 0.880 $\pm$ 0.009	 &	 0.075 $\pm$ 0.003	 &	 0.203 $\pm$ 0.008	 &	N	 \\ 
4.95	 &	  0.442 $\pm$ 0.002	 &	 0.941 $\pm$ 0.005	 &	 0.221 $\pm$ 0.012	 &	 0.541 $\pm$ 0.045	 &	SmA 	 \\ 
5.50	 &	  0.471 $\pm$ 0.003	 &	 0.963 $\pm$ 0.003	 &	 0.342 $\pm$ 0.015	 &	 0.708 $\pm$ 0.032	 &	SmA 	 \\ 
6.05	 &	  0.495 $\pm$ 0.004	 &	 0.971 $\pm$ 0.001	 &	 0.442 $\pm$ 0.013	 &	 0.793 $\pm$ 0.016	 &	SmA 	 \\ 
6.60	 &	  0.517 $\pm$ 0.002	 &	 0.977 $\pm$ 0.002	 &	 0.505 $\pm$ 0.018	 &	 0.837 $\pm$ 0.061	 &	SmA 	 \\ 
7.15	 &	  0.532 $\pm$ 0.004	 &	 0.981 $\pm$ 0.001	 &	 0.531 $\pm$ 0.013	 &	 0.850 $\pm$ 0.026	 &	SmA 	 \\ 
7.70	 &	  0.546 $\pm$ 0.002	 &	 0.984 $\pm$ 0.001	 &	 0.542 $\pm$ 0.004	 &	 0.861 $\pm$ 0.050	 &	SmA 	 \\ 
8.25	 &	  0.560 $\pm$ 0.003	 &	 0.986 $\pm$ 0.002	 &	 0.554 $\pm$ 0.004	 &	 0.866 $\pm$ 0.033	 &	SmA 	 \\ 
9.90	 &	  0.621 $\pm$ 0.003	 &	 0.996 $\pm$ 0.000	 &	 0.664 $\pm$ 0.005	 &	 0.899 $\pm$ 0.003	 &	X	 \\ 
10.45	 &	  0.629 $\pm$ 0.004	 &	 0.996 $\pm$ 0.001	 &	 0.677 $\pm$ 0.010	 &	 0.902 $\pm$ 0.012	 &	X	 \\ 
11.00	 &	  0.637 $\pm$ 0.001	 &	 0.996 $\pm$ 0.001	 &	 0.679 $\pm$ 0.003	 &	 0.915 $\pm$ 0.011	 &	X	 \\ 
	\hline
	\end{tabular}
\end{table}

     \begin{table}[h!]
	\caption{$L/D= 10.00 $}
	\centering
        \footnotesize
\label{ld10}
	 \begin{tabular}{c c c c c c}
	\hline
	$P^*$ &  	 $\eta$ &  	  $S$ &   	 $\psi_6$ &	 $\tau$ & 	 Phase \\ \hline
0.79	 &	  0.165 $\pm$ 0.001	 &	 0.024 $\pm$ 0.008	 &	 0.000 $\pm$ 0.001	 &	 0.367 $\pm$ 0.161	 &	I	 \\ 
1.57	 &	  0.224 $\pm$ 0.001	 &	 0.043 $\pm$ 0.013	 &	 0.002 $\pm$ 0.001	 &	 0.273 $\pm$ 0.366	 &	I	 \\ 
2.36	 &	  0.299 $\pm$ 0.042	 &	 0.804 $\pm$ 0.113	 &	 0.007 $\pm$ 0.001	 &	 0.220 $\pm$ 0.004	 &	N	 \\ 
3.93	 &	  0.386 $\pm$ 0.002	 &	 0.953 $\pm$ 0.003	 &	 0.033 $\pm$ 0.004	 &	 0.217 $\pm$ 0.004	 &	N	 \\ 
4.71	 &	  0.432 $\pm$ 0.002	 &	 0.971 $\pm$ 0.002	 &	 0.118 $\pm$ 0.008	 &	 0.543 $\pm$ 0.022	 &	SmA 	 \\ 
5.50	 &	  0.475 $\pm$ 0.001	 &	 0.982 $\pm$ 0.002	 &	 0.270 $\pm$ 0.016	 &	 0.763 $\pm$ 0.010	 &	SmA 	 \\ 
6.28	 &	  0.504 $\pm$ 0.002	 &	 0.986 $\pm$ 0.001	 &	 0.374 $\pm$ 0.007	 &	 0.826 $\pm$ 0.019	 &	SmA 	 \\ 
7.07	 &	  0.528 $\pm$ 0.002	 &	 0.989 $\pm$ 0.001	 &	 0.451 $\pm$ 0.014	 &	 0.868 $\pm$ 0.023	 &	SmA 	 \\ 
7.85	 &	  0.548 $\pm$ 0.002	 &	 0.991 $\pm$ 0.001	 &	 0.509 $\pm$ 0.014	 &	 0.879 $\pm$ 0.014	 &	SmA 	 \\ 
8.64	 &	  0.570 $\pm$ 0.002	 &	 0.993 $\pm$ 0.001	 &	 0.543 $\pm$ 0.010	 &	 0.901 $\pm$ 0.018	 &	SmA 	 \\ 
9.42	 &	  0.611 $\pm$ 0.001	 &	 0.998 $\pm$ 0.001	 &	 0.619 $\pm$ 0.007	 &	 0.880 $\pm$ 0.004	 &	X	 \\ 
10.21	 &	  0.628 $\pm$ 0.002	 &	 0.998 $\pm$ 0.001	 &	 0.670 $\pm$ 0.004	 &	 0.900 $\pm$ 0.014	 &	X	 \\ 
11.00	 &	  0.642 $\pm$ 0.001	 &	 0.998 $\pm$ 0.001	 &	 0.704 $\pm$ 0.007	 &	 0.905 $\pm$ 0.017	 &	X	 \\ 
11.78	 &	  0.652 $\pm$ 0.003	 &	 0.999 $\pm$ 0.001	 &	 0.710 $\pm$ 0.011	 &	 0.912 $\pm$ 0.011	 &	X	 \\ 
	\hline
	\end{tabular}
\end{table}

\clearpage
%

\end{document}